\documentclass[12pt]{article}
\pdfoutput=1
\newcommand{\blind}{0}
\newcommand{\temptitle}{
	Bias correction in daily maximum and minimum temperature measurements through Gaussian process modeling
}
\usepackage{mathrsfs}
\usepackage{graphicx}
\usepackage{xr-hyper} 
\usepackage{xcolor} 
\usepackage{geometry} 
\usepackage{amsmath} 
\usepackage{amssymb} 
\usepackage{hyperref}
\hypersetup{
    colorlinks=true,
    pdfborder=0 0 1,
    pdftitle=\temptitle{},
    allcolors=[rgb]{0.00,0.38,0.59},
    allbordercolors=[rgb]{0.60,0.76,0.86}
}
\usepackage[title,toc]{appendix}
\usepackage{natbib}
\usepackage{bm}
\usepackage{mathtools}
\usepackage{pdflscape}
\usepackage[singlespacing,nodisplayskipstretch]{setspace}
\makeatletter
\renewcommand*\env@cases[1][0.6]{%
  \let\@ifnextchar\new@ifnextchar
  \left\lbrace
  \def\arraystretch{#1}%
  \array{@{}l@{\quad}l@{}}%
}
\makeatother
\usepackage{authblk}

\usepackage{enumitem}
\newlist{flatlist}{enumerate*}{1}
\setlist[flatlist]{label=(\arabic*)}

\def\equationautorefname~#1\null{(#1)\null}

\newcommand{\keywords}{Gaussian processes; kriging; Climate; Bayesian computation; }
\hypersetup{pdfkeywords=\keywords{}}
\hypersetup{pdfauthor=Maxime Rischard}
\newcommand{\tempauthors}{
\author[1]{Maxime Rischard\thanks{We thank Peter Huybers, Debdeep Pati and Martin Lysy for their ideas, questions and suggestions.}}
\author[2]{Karen A. McKinnon}
\author[1]{Natesh Pillai}
\affil[1]{Department of Statistics, Harvard University}
\affil[2]{National Center for Atmospheric Research; Descartes Labs}

}

\newcommand{\mathbold}[1]{\bm{#1}} 
\DeclarePairedDelimiter{\parenthesis}{\lparen}{\rparen}
\DeclarePairedDelimiter{\squarebracket}{\lbrack}{\rbrack}
\DeclarePairedDelimiter{\curlybracket}{\lbrace}{\rbrace}

\newcommand{\del}[1]{\parenthesis*{#1}}
\newcommand{\sbr}[1]{\squarebracket*{#1}}
\newcommand{\cbr}[1]{\curlybracket*{#1}}

\DeclareMathOperator*{\argmax}{arg\,max}

\let\Pr\relax
\DeclareMathOperator{\Pr}{\mathbb{P}}
\DeclareMathOperator{\E}{\mathbb{E}}

\DeclareMathOperator{\cov}{{cov}}
\DeclareMathOperator{\var}{{var}}
\DeclareMathOperator{\mse}{{MSE}}
\DeclareMathOperator{\Ind}{\mathbb{I}}

\DeclareMathOperator{\normal}{\mathcal{N}}

\DeclareMathOperator{\GP}{\mathcal{GP}}
\newcommand{\T}{\mathrm{T}}
\newcommand{\Tn}{\T_{n}}
\newcommand{\Tx}{\T_{x}}
\newcommand{\station}[1]{\mathrm{station}\sbr{#1}}
\newcommand{\xvec}{\mathbold{x}}
\newcommand{\hvec}{\mathbold{h}}
\newcommand{\indep}{\perp}
\newcommand{\iid}{iid}
\newcommand{\trans}{^{\intercal}}

\newcommand{\sigman}{\sigma_{\epsilon}}
\newcommand{\degreeC}{{}^{\circ}\mathrm{C}}
\newcommand{\miss}{\mathrm{miss}}
\newcommand{\obs}{\mathrm{nearby}}
\newcommand{\error}{\mathrm{err}}
\newcommand{\hour}{\mathtt{hr}}
\DeclareMathOperator*{\softmax}{smoothmax}
\DeclareMathOperator*{\softmin}{smoothmin}

\DeclareMathOperator{\kSESE}{k_{\mathtt{SExSE}}}
\DeclareMathOperator{\kdiurn}{k_{\mathtt{SESE_{24}}}}
\DeclareMathOperator{\ksumprod}{k_{\mathtt{sumprod}}}
\newcommand{\iday}{d}
\newcommand{\tmeas}{t^{\mathrm{meas}}}
\newcommand{\dayset}[1]{(\tmeas_{#1-1},\,\tmeas_{#1}]}
\newcommand{\concordance}{\delta}
\newcommand{\Xmax}{X_{\max}}
\newcommand{\Xmin}{X_{\min}}
\newcommand{\Fcond}{F_{X \mid \Xmax,\Xmin}}
\newcommand{\pxx}[2]{\Pr{}_{#1#2}}
\newcommand{\pij}{\pxx{i}{j}}
\newcommand{\pisum}{\pxx{i}{\bullet}}
\newcommand{\psumj}{\pxx{\bullet}{j}}

\newcommand{\subspace}{\mathrm{space}}
\newcommand{\subtime}{\mathrm{time}}

\newcommand{\Smm}{\mathbf{\Sigma}_{\miss,\miss}}
\newcommand{\Kmo}{\mathbf{K}_{\miss,\obs}}

\newcommand{\Soo}{\mathbold{\Sigma}_{\obs,\obs}}

\begin{document}


\if0\blind
{
\title{
    \Large
    \bf
    \temptitle
}
\tempauthors{}
\maketitle
} \fi

\if1\blind
{
  \bigskip
  \bigskip
  \bigskip
  \begin{center}
    {\LARGE\bf \temptitle}
\end{center}
  \medskip
} \fi

\begin{abstract}
The Global Historical Climatology Network-Daily database contains, among other variables, daily maximum and minimum temperatures from weather stations around the globe.
It is long known that climatological summary statistics based on daily temperature minima and maxima will not be accurate, if the bias due to the time at which the observations were collected is not accounted for. Despite some previous work, to our knowledge, there does not exist a satisfactory solution to this important problem.
In this paper, we carefully detail the problem and develop a novel approach to address it. Our idea is to impute the hourly temperatures
at the location of the measurements by borrowing information from the nearby stations that record hourly temperatures, which then can be used to create accurate summaries of temperature extremes.
The key difficulty is that these imputations of the temperature curves must satisfy the constraint of falling between the observed daily minima and maxima, and attaining those values at least once in a twenty-four hour period.
We develop a spatiotemporal Gaussian process model for imputing the hourly measurements from the nearby stations, and then develop a novel and easy to implement Markov Chain Monte Carlo technique to sample from the posterior distribution satisfying the above constraints. 
We validate our imputation model using hourly temperature data 
from four meteorological stations in Iowa,
of which one is hidden and the data replaced with daily minima and maxima,
and show that the imputed temperatures recover the hidden temperatures well.
We also demonstrate that our model can exploit information contained in the data to infer the time of daily measurements. 
\end{abstract}

\section{Introduction}\label{sec:introduction}

Long, high-quality records of temperature provide an important basis for our understanding of climate variability and change. Historically, there has been a focus on monthly-average temperature records that are sufficient for certain analyses, such as quantifying long-term changes in temperature. As our knowledge of climate change expands, however, there is increasing interest in understanding changes in temperature on shorter timescales, with a particular focus on extreme events. To do so, it is necessary to utilize temperature data with higher temporal resolution. 

Recent work has led to the development of the Global Historical Climatology Network-Daily (GHCND) database \citep{menne2012overview}, which contains, among other variables, daily maximum and minimum temperatures from weather stations around the globe. The database draws from a range of different sources, and the data within it undergoes basic quality control to remove erroneous values. 

The current quality control methodology, however, does not account for so-called ``inhomogeneities.'' Inhomogeneities result from changes in measurement practices that impact the recorded temperatures. For temperature, known inhomogeneities include 
\begin{flatlist}
    \item changes in the time of observation, 
    \item changes in the thermometer technology, 
    \item station relocation, and 
    \item changes in land use around a station \citep{menne2009us}.
\end{flatlist}
While these inhomogeneities have a small effect on, for example, the estimation of global mean temperature, they can have a large effect on estimation of temperature variability and change at a more local scale.

There is a large body of work focused on homogenizing monthly-average temperatures \citep[e.g.,][]{karl1986model, easterling1996development, peterson1998homogeneity, ducre2003comparison, menne2009homogenization, vincent2012second}, resulting in widely available, large-scale homogenized monthly temperature datasets. 
Homogenization typically proceeds through identifying non-climatic `breakpoints' in a given time series through comparison with neighboring stations.
Once a breakpoint is identified, the measurements recorded after the breakpoint are adjusted in some way to reduce or remove the inhomogeneity.
Most applications of these methods, however, focus on adjusting the mean state of the data rather than the shape of the distribution \citep[see][and references therein]{della2006method}.
While this may be sufficient for monthly data, it is known that changes in measurement practices may affect different quantiles of the daily temperature distribution unequally.
To address this issue, some homogenization methods have also employed frequency distribution matching techniques, so that each temperature recorded after a breakpoint is adjusted according to its percentile within the time series \citep{della2006method, trewin2013daily}. 

\subsection{The Problem} 
\label{sec:theproblem}
\label{sec:illustrate_bias}

\begin{figure}[tbp]
\centering
\includegraphics[height=0.4\textheight,width=0.99\textwidth,keepaspectratio]{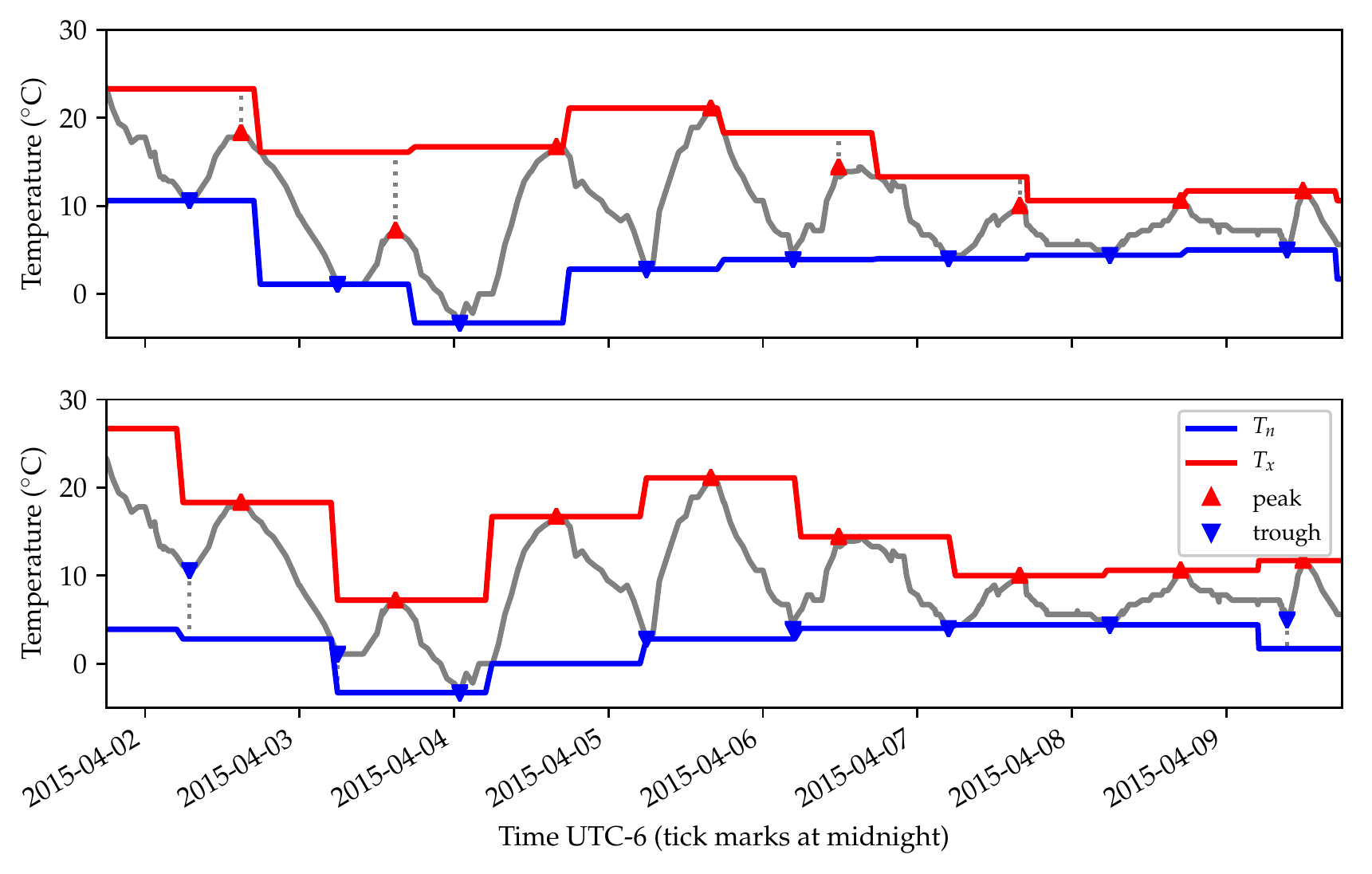}
\caption{
\label{fig:waterloo_triangles}
An extract of the temperature measurements from KALO.
The blue and red triangles respectively indicate the coldest and warmest temperature of each diurnal cycle.
The blue and red lines respectively show the observed maximum and temperature recorded each day at 17:00 (top) or 5:00 (bottom) for the 24-hour period preceding the measurement.
Discrepancies between the 24-hour extrema, and the peaks and troughs of the diurnal cycle, are indicated with dotted lines.
}
\end{figure}

Many historical measurements of daily temperatures are provided as daily maximum and minimum temperatures (\(\Tx\) and \(\Tn\) respectively), which ideally measure the peak and trough of each diurnal temperature cycle. 
\(\Tx\) and \(\Tn\) are often recorded by an observer who every 24 hours visits a weather station equipped with a maximum-minimum thermometer, and notes the maximum and minimum registered by the instrument in the last 24 hours.
In this section we explain how this measurement practice can cause the \(\Tx\) and \(\Tn\) measurements to fail to capture the peaks and troughs of some diurnal cycles.
This has long been recognized in the scientific literature; see for example \citet{baker1975effect} and references therein.

\begin{figure}[tbp]
\centering
\includegraphics[width=0.6\textwidth]{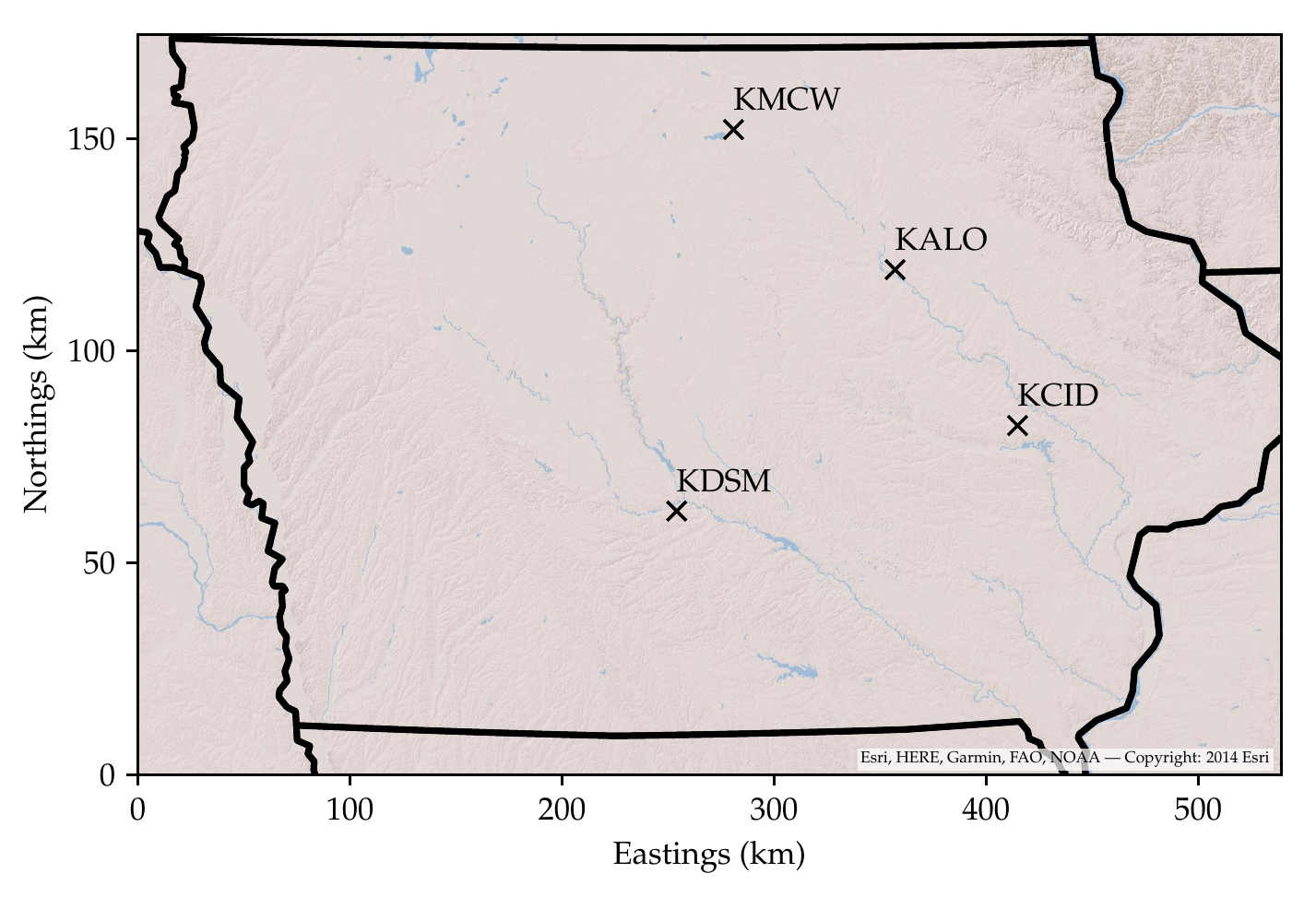}
\caption{
\label{fig:Iowa_map}
Map of the four airport weather stations in Iowa providing hourly temperature records. 
Each airport is identified by its ICAO code.
}
\end{figure}

\autoref{fig:waterloo_triangles} illustrates the problem with ten days of hourly temperature measurements from the Waterloo Municipal Airport (KALO) weather station in Iowa. 
\autoref{fig:Iowa_map} gives a map of the four Iowa weather stations used as examples throughout this paper.
We emulate daily \(\Tx\)/\(\Tn\) measurements by dividing the data into 24 hour measurement windows, and reporting the minimum and maximum temperature that was recorded in this window.
On most days, the measurements successfully capture the peak and trough of the diurnal cycle.
But there are also several discrepancies (indicated with vertical dotted lines), typically in \(\Tx\) when the measurements are made near the warmest hour of the day, and in \(\Tn\) when the measurements are made near the coldest hour.
A blatant example occurs on April 3rd,
where the peak of the diurnal cycle is 7.2\(\degreeC\) and occurs at 15:00 (all times are in the UTC-6 time zone, and tick marks are at midnight at the start of each day), but with measurements made at 17:00, the day's \(\Tx\) record of 16.1\(\degreeC\) is reached immediately after the previous day's measurement: a 8.9\(\degreeC\) overestimate.
Ideally, measurements of the diurnal cycle peak and trough would be obtained by recording \(\Tx\) and \(\Tn\) at the coldest and warmest time of day respectively.
This would minimize the possibility of the previous or next diurnal cycle setting the measured \(\Tx\) or \(\Tn\).
For convenience, however, most observers instead record data at a single daytime hour.
Our goal is to address the bias that results from this measurement practice.

\begin{figure}[tbp]
\centering
\includegraphics[height=0.3\textheight,width=0.99\textwidth,keepaspectratio]{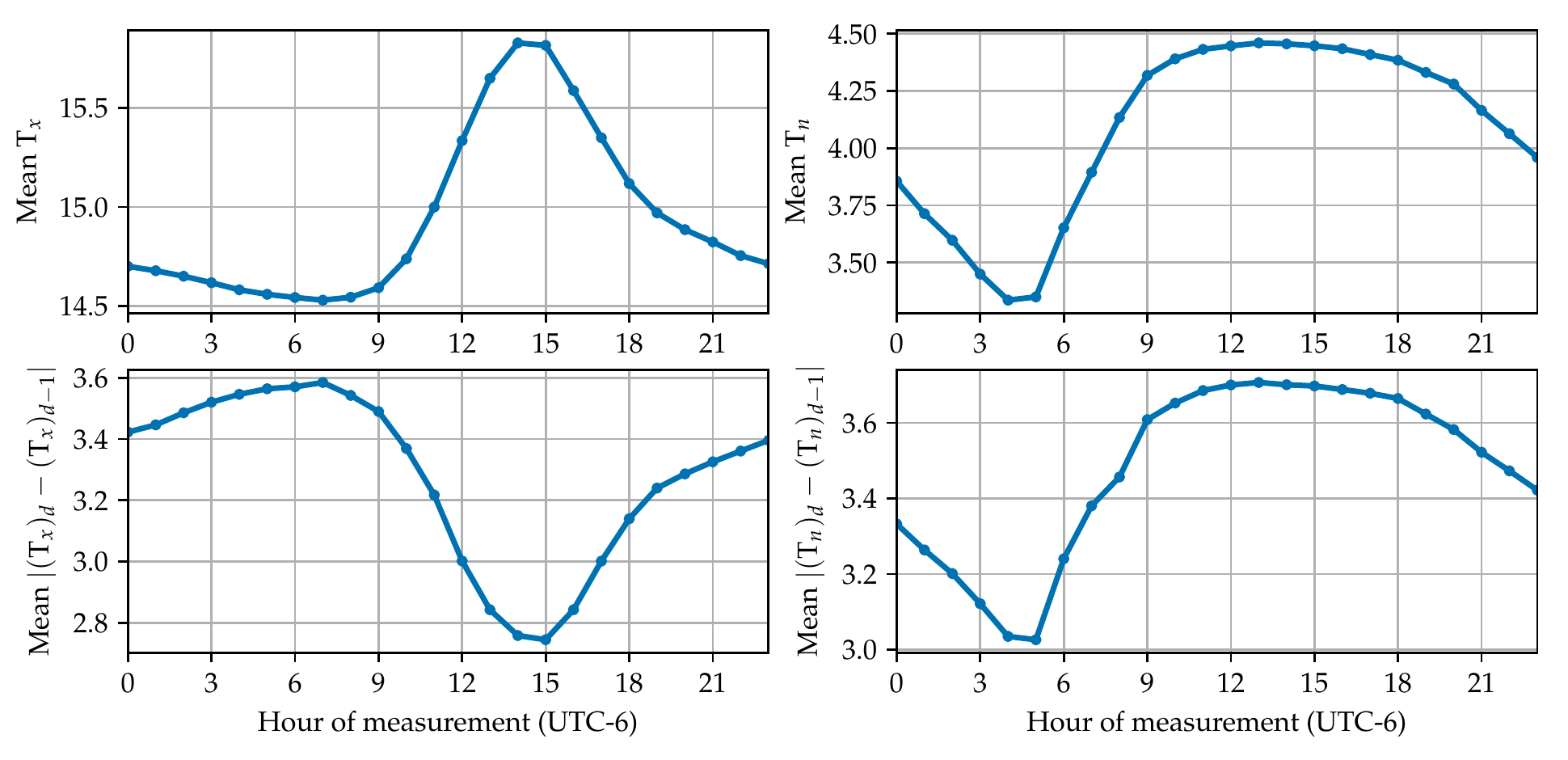}
\caption{
\label{fig:waterloo_avgTnTx}
Mean daily \(\Tx\) (top left) and \(\Tn\) (top right), and
mean absolute daily change in \(\Tx\) (bottom left) and \(\Tn\) (bottom right),
extracted from hourly temperature records at KALO in 2015,
under varying measurement hours of \(\Tx\) and \(\Tn\).}
\end{figure}

The bias in the daily records can in turn induce bias in the long-term summary statistics that are of climatological interest.
A statistic as simple as the average daily maximum temperature for an entire year (2015) increases by over 1\(\degreeC\) if the measurements are made at 15:00 compared to 9:00, as seen in \autoref{fig:waterloo_avgTnTx}.
Conversely, the average \(\Tn\) is colder by over 1\(\degreeC\) if \(\Tn\) is measured at 5:00 rather than 15:00.

If the time of observation remained constant over time, this systematic bias would still exist, but it would not be linked to spurious trends in the data. However, there have been known (and likely unknown) changes in the time of observation.
In the United States, for example, observers were instructed to switch from recording data in the afternoon to recording data in the morning beginning in the 1950s.
This change has led to an apparent decrease in both \(\Tx\) and \(\Tn\) over time \citep{menne2009us}. 
Such spurious trends also compromise the study of weather variability, through summary statistics such as the average absolute change in daily temperature maxima and minima from one day to the next, as seen in \autoref{fig:waterloo_avgTnTx}.



\subsection{Our Approach}\label{sec:approach}
    
One of our goals is to be able to infer the ``true'' \(\Tx\) and \(\Tn\) peaks and troughs of the diurnal cycle throughout the data records, so as to correct both the variance biases and the spurious trends.
This stands in constrast to previous work, which has focused directly on addressing spurious trends.
We approach the problem as a missing data problem: 
if we had access to the full temperature time series at the station rather than just \(\Tx\) and \(\Tn\) measured at an arbitrary time, 
we would be able to retrospectively choose the hour of measurements, to avoid the issues described in \autoref{sec:theproblem}.
Our idea therefore is to impute the hourly time series of temperatures
at the location of the \(\Tx\)/\(\Tn\) measurements.
In turn, the imputed time series can be used to create accurate summaries of temperature extremes.

Our imputation strategy is to borrow information from the nearby weather stations, usually located at airports, that record the current temperature about once an hour. 
Although it should be noted that the sampling times are not always equally spaced, we refer to these measurements as ``hourly'' throughout this paper.
They cannot be used directly for climatology, as the weather stations that provide them are not always as carefully documented, calibrated, and situated as the research stations included in the GHCND. 
For instance, weather stations at locations experiencing a lot of human activity, like airports, may record higher temperatures on average.
However, even if mis-calibrated or systematically biased, the time series data from these nearby stations do contain valuable information about the hourly changes in temperatures on any given day.

In this paper, we develop a spatiotemporal Gaussian process model pooling the information from nearby stations with hourly data and simulate multiple realizations of hourly temperature time series at each station of interest.
The key technical difficulty is that these imputations of the temperature curves must satisfy the constraint of falling between the observed daily minima and maxima, and attaining those values at least once in a twenty-four hour period.
We develop SmoothHMC, a novel and easy to implement Markov Chain Monte Carlo (MCMC) algorithm to sample from the posterior distribution satisfying the above constraints.
Our constrained imputations are implemented in the Stan programming language \citep{stancite}; our code is publicly available on the first author's GitHub account.
Compared to a custom implementation, the Stan model code is short and Stan's MCMC samplers are well-optimized, which makes our imputation strategy efficient and easy to reproduce.

\section{A First Spatiotemporal Model}\label{a-spatiotemporal-model}

In order to pool the information from temperatures measured at various locations and times, we develop a spatio-temporal Gaussian process model.
In its simplest form, we posit that temperatures from stations that are near each other are more correlated than distant stations, and that those correlations also decay in time.
We model the simultaneous temperatures throughout a region as a Gaussian process, with covariance between two locations \(\xvec\) and \(\xvec'\) given by the squared exponential (SE) covariance with characteristic lengthscale \(\ell_{\subspace}\) and variance \(\sigma_{\subspace}^2\):
\begin{equation}
    \cov\del{T(\xvec), T(\xvec') \mid t} = k_{\subspace}(\xvec, \xvec') =  \sigma_{\subspace}^2 \exp\del{-\frac{\del{\xvec-\xvec'}\trans\del{\xvec-\xvec'}}{2\ell_{\subspace}^2}}\,.
    \label{eq:kspace}
\end{equation}
Similarly, the time series of temperatures at a single location can be modeled as a Gaussian process with characteristic timescale \(\ell_{\subtime}\) and variance \(\sigma_{\subtime}^2\):
\begin{equation}
\cov\del{T(t), T(t') \mid \xvec} = k_{\subtime}(t, t') = \sigma_{\subtime}^2 \exp\del{-\frac{\del{t-t'}^2}{2\ell_{\subtime}^2}}\,.
\end{equation}
We combine the spatial and temporal model by multiplying the covariances functions:
\begin{equation}
k(\xvec,\xvec',t,t') = k_{\subtime}(t,t') \cdot k_{\subspace}(\xvec, \xvec')\,.
\end{equation}
This yields the covariance of the Gaussian process underlying the spatio-temporal model of temperatures.
The variances \(\sigma_{\subspace}^2\) and \(\sigma_{\subtime}^2\) are not separately identifiable, so we arbitrarily fix \(\sigma_{\subspace}^2=1\).
To allow for systematic differences between stations, we add a mean temperature parameter \(\mu_{\station{i}}\) for each station, where \(\station{i}\) is the index of the station at which observation \(i\) was recorded.
This parameter captures both systematic differences in temperature between locations, for example due to differences in altitude, vegetation, or built environment around the station, and also calibration errors in the measurement apparatus.

The observation model depends on the type of measurement obtained at a given location.
At stations \(j\) that provide a full temperature time series, we model the \(i^{\mathrm{th}}\) temperature record as a noisy measurement from the true time series, with iid normal noise:
\begin{equation}
\begin{split}
    & \T_{ij} = \mu_{j} + f(\xvec_j, t_{ij}) + \epsilon_{ij}\,,\quad
    \epsilon_{ij} \overset{\iid}{\sim} \normal\del{0,\sigman^2}\,\\
    & f(\xvec_j, t_{ij}) \sim \GP\del{0, k_{}(\xvec,\xvec',t,t')}\,.\\
\end{split}
\label{eq:gpmodel}
\end{equation}
The noise term captures measurement error and micro-fluctuations occuring on time scales much shorter than \(\ell_{\subtime}\).
At stations \(j\) that only provide daily \(\Tx\) and \(\Tn\) records, 
we denote the time of the \(\iday^{\mathrm{th}}\) daily measurement by \(\tmeas_{\iday}\), 
and approximate the \(\Tx\) and \(\Tn\) observation respectively as the maximum or minimum temperatures at a discretized set of times \(t_{ij}\) inside of \(\dayset{\iday}\):
\begin{equation}
\begin{split}
    \del{\Tx}_{\iday j} &= \max\cbr{\T_{ij}\text{, for all \(i\) such that } t_{ij} \in \dayset{\iday}} \,,\\
    \del{\Tn}_{\iday j} &= \min\cbr{\T_{ij}\text{, for all \(i\) such that } t_{ij} \in \dayset{\iday}} \,,
\end{split}
\label{eq:obs_Tn_Tx}
\end{equation}
with \(\T_{ij}\) modeled as in \autoref{eq:gpmodel}.

\subsection{Fitting the Spatiotemporal Model}\label{fitting-the-spatiotemporal-model}

Software is readily available in many programming languages for fitting Gaussian process models, including inference on the covariance parameters. We chose to use the julia \texttt{GaussianProcesses.jl} package to fit the above spatiotemporal model to the hourly temperatures at four Iowa weather stations.
The Iowa data set includes 47,864 measurements, which is computationally challenging to fit directly with a single Gaussian process.
There are many methods to handle large data sets with Gaussian processes: for example \citet{quinonero2007approximation} review sparse approximations to Gaussian processes from a machine learning perspective, while \citet{banerjee2008gaussian} develop a method specifically for large spatial data sets.
For simplicity, we chose instead to divide the data into 10-day chunks, modeled as independent Gaussian processes with shared hyperparameters.
We put weak normal priors on \(\mu_{\station{i}}\) with large standard deviation \(\sigma_{\mu}=10\,\degreeC\), which can be incorporated into the Gaussian process with an additional term
\begin{equation}
    k_{\mu}(\xvec, \xvec') = \begin{cases}
\sigma_\mu^2 &\text{if } \xvec = \xvec'\,, \\
0 &\text{otherwise}\,.
\end{cases}
\end{equation}
added to the covariance function.
The spatio-temporal covariance function becomes
\begin{equation}
    \kSESE(\xvec,\xvec',t,t') = k_{\subtime}(t,t') \cdot k_{\subspace}(\xvec, \xvec') + k_\mu(\xvec, \xvec') \,,
    \label{eq:ksese}
\end{equation}
which we denote \(\kSESE\) to distinguish it from the covariance functions developed later in \autoref{sec:improving_model}.
Our model thus has four free parameters, \(\sigma_{\subtime}\), \(\ell_{\subtime}\), \(\ell_{\subspace}\) and \(\sigman\), which we fit by maximizing the marginal likelihood of \(\T\), the complete 2015 temperature time series provided at the four Iowa weather stations:
\begin{equation}
\label{eq:optimization}
\widehat\sigma_{\subtime},\widehat\ell_{\subtime},\widehat\ell_{\subspace},\widehat\sigman = \argmax_{\sigma_{\subtime},\ell_{\subtime},\ell_{\subspace},\sigman} \cbr{ \Pr\del{ \T \mid {\sigma_{\subtime},\ell_{\subtime},\ell_{\subspace},\sigman} } }\,.
\end{equation}
The fitted covariance  values are found in \autoref{table:fitted_params}.

\begin{table}[tbp]
\begin{center}
\bgroup
\def\arraystretch{1.1}
\addtolength{\tabcolsep}{6pt}  
\begin{tabular}{r@{ }lrrr}
\hline
&& \multicolumn{3}{c}{Covariance Function} \\
\multicolumn{2}{l}{Parameter} & \multicolumn{1}{l}{\(\kSESE\)} & \multicolumn{1}{l}{\(\kdiurn\)} & \multicolumn{1}{l}{\(\ksumprod\)} \\
\hline
$\sigman$ &	($\degreeC$) &	0.4 &	0.4 &	0.2\\ 
\hline
$\sigma_\subtime$ &	($\degreeC$) &	3.7 &	3.1 &	0.5,\,0.9,\,4.4\\ 
$\ell_\subtime$ &	($\mathrm{hr}$) &	2.7 &	2.8 &	0.3,\,1.9,\,8.9\\ 
$\ell_\subspace$ &	($\mathrm{km}$) &	176 &	154 &	10,\,59,\,370\\ 
$\alpha_{\subtime}$ &	 &		 &		 &	0.3,\,1.1,\,0.3\\ 
\hline
$\sigma_{24}$ &	($\degreeC$) &		 &	2.4 &	2.7\\ 
$\ell_{24}$ &	($\mathrm{hr}$) &		 &	0.7 &	0.8\\ 
$\ell_{\subspace{}24}$ &	($\mathrm{km}$) &		 &	1414 &	785\\ 
\hline
\end{tabular}
\addtolength{\tabcolsep}{-3pt}  
\caption{
    Fitted parameters for each specification of the Gaussian process covariance function.
    For \(\ksumprod\) \autoref{eq:sumprod_kernel}, the parameters of the short-term, medium-term, and long-term components are separated by commas.
    Notice how shorter timescales \(\ell_\subtime\) are associated with shorter lengthscales \(\ell_\subspace\) by the fitted covariance function.
    \label{table:fitted_params}
}
\egroup
\end{center}
\end{table}
    
\section{Predictions Using Nearby Data}\label{predictions-using-nearby-data}

\label{sec:predict_nearby}
\begin{figure}[tb]
\centering
\includegraphics[width=0.99\textwidth]{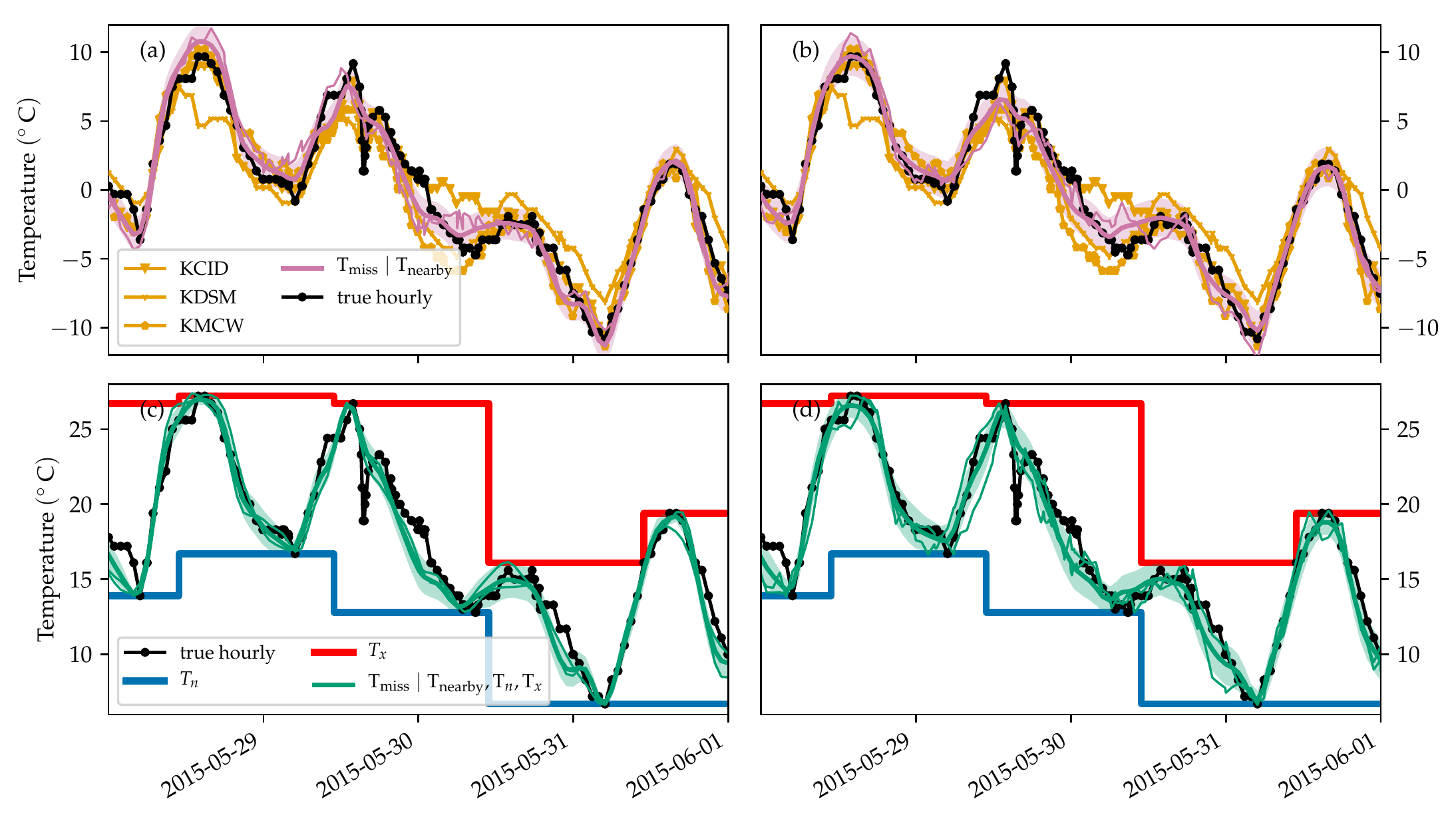}
\caption{\label{fig:imputations_2x2}Imputations of the temperature time series at Waterloo Municipal Airport (KALO) between May 28, 2015 and June 1, 2015 (a) using only nearby data and the product of squared exponentials model; (b) using only nearby data and the sum of products model; (c) incorporating \(\Tn\) and \(\Tx\) measurements under the product of squared exponentials model; and (d) incorporating \(\Tn\) and \(\Tx\) measurements under the sum of products model. The mean is subtracted from each time series in (a) and (b) as the models leave the average temperature at the imputation site as a free parameter. For each imputation distribution, the mean is shown as a thick line, surrounded by an 80\% credible envelope in lighter color, and example imputations as thinner lines.}
\end{figure}

After optimizing the parameters of the spatio-temporal covariance \autoref{eq:ksese}, we use the model \autoref{eq:gpmodel}---fitted to the data from nearby stations with full time series---to provide time series predictions at the station that only collects \(\Tx\) and \(\Tn\) data.
Gaussian processes give closed-form expressions for the posterior distribution of the predicted temperatures.
We denote the temperatures we wish to impute as \(\T_\miss{}\) at times \(t_\miss\) and location \(\xvec_\miss\) and those observed at nearby stations as \(\T_\obs{}\), at times \(t_\obs\) and locations \(X_\obs\).
Under the spatio-temporal model \autoref{eq:gpmodel}, \(\T_\miss\) and \(\T_\obs\) are jointly multivariate normal, with mean zero and covariance given by \(\kSESE(\xvec,\xvec',t,t')\).
Standard results for conditioning within multivariate normals then yield:
\begin{equation}
\begin{split}
    \T_\miss \mid \T_\obs &\sim \normal\del{\mu_{\miss \mid \obs}, \Sigma_{\miss \mid \obs}}\,\text{, with} \\
    \mu_{\miss \mid \obs} &= \E \del{\T_\miss \mid \T_\obs} = \Kmo \Soo^{-1} \T_\obs \,\text{, and} \\
    \Sigma_{\miss \mid \obs} &= \var \del{\T_\miss \mid \T_\obs} = \Smm - \Kmo \Soo^{-1} \Kmo\trans
    \,.
\end{split}
\label{eq:unconstrained_post}
\end{equation}
All covariance matrices can be derived from the model.
For example, the \(ij^{\text{th}}\) entry of \(\Kmo = \cov\del{\T_\miss, \T_\obs}\) is given by \(\kSESE(\xvec_\miss,X_\obs\sbr{j},t_\miss\sbr{i},t_\obs\sbr{j})\), where \(X_\obs\sbr{j}\) gives the location of the \(j\)th observation, and \(t_\obs\sbr{j}\) its time.
The two \(\mathbold{\Sigma}\) matrices have an additional \(\sigman^2\) diagonal component for measurement noise.

In \autoref{fig:imputations_2x2}(a), we show an example of predictions obtained from this spatio-temporal model. 
We withheld temperature measurements from KALO (shown in black), and then used data from the three remaining stations (KCID, KDSM and KMCW, shown in orange) to predict the 2015 temperature time series At KALO.
To speed up computations, we process 73 days of data at a time, with 48 days overlapping between adjacent prediction windows so that predictions can always be made away from the edge of the prediction window (except at the start and end of the year).
The predictions can be seen to combine information from the three other stations, giving less weight to KDSM, which is further away from KALO.
We will discuss the quality of these predictions in more detail in \autoref{sec:diagnostics}, after completing the exposition of our imputation strategy.

\section{Imputing by Conditioning on Extrema}
\label{imputations}
    
Our aim is not simply to predict temperatures at a location with no measurements, but rather to impute hourly temperatures at a location with accurate measurements of the daily temperature extrema.
This is an instance of a more general statistical problem: if a random \(p\)-vector \(\cbr{X_i:~i=1,\dotsc,p}\) has a known distribution \(F_X\), and its maximum \(\Xmax = \max_i\cbr{X_i}\) and minimum \(\Xmin = \min_i\cbr{X_i}\) are measured, how does one draw samples from \(\Fcond\), the distribution of \(X\) conditional on \(\Xmax\) and \(\Xmin\)?
Conditional draws from \(\Fcond\) need to respect three constraints: one component of \(X\) must be equal to \(\Xmin\), another to \(\Xmax\), and all other components must lie between \(\Xmin\) and \(\Xmax\).

Conceptually, we could implement a valid imputation algorithm by drawing random samples \(F_X\),
and accepting only those samples that satisfy the three constraints.
Unfortunately, if \(F_X\) is a continuous distribution, the probability of a random draw from \(F_X\) satisfying such sharp constraints is zero.
One could envision adding some tolerance, so that samples with minimum and maximum within a small margin of \(\Xmax\) and \(\Xmin\) are retained, but as the dimensionality \(p\) grows, the rejection probability will rapidly go to 1, thus requiring huge sample sizes.
Ultimately, this rejection sampling strategy is therefore bound to fail.

Markov Chain Monte Carlo (MCMC) techniques can also be used to draw samples from arbitrary distributions with densities known up to a constant. The density of \(\Fcond\) is obtained up to a constant multiplier through a simple application of Bayes' theorem. It is proportional to the prior density of \(F_X\) multiplied by indicators ensuring that the extrema are respected:
\begin{equation}
    \Pr\del{X \mid \Xmax,\Xmin} 
        \propto \Pr\del{X} \Ind\cbr{ \max_i\cbr{X_i} = \Xmax }\Ind\cbr{ \min_i\cbr{X_i} = \Xmin } \,.
\label{eq:bayes_exact}
\end{equation}
However, once again, this distribution is zero everywhere in \(\mathbb{R}^p\), except in a (p-2) dimensional subspace where the \(\min\) and \(\max\) constraints are met.
Consequently, out-of-the-box generic MCMC algorithms targeting \autoref{eq:bayes_exact} will not successfully converge to \(\Fcond\).
We therefore loosen the constraint by replacing the likelihood term \(\Pr\del{\Xmax,\Xmin \mid X}\) with two narrow independent normal distributions around the minimum and maximum of \(X\):
\begin{equation}
\begin{split}
    \Pr\del{X \mid \Xmax,\Xmin} &\propto \Pr\del{X} 
                                         \normal\del{\Xmax \mid \max_i\cbr{X_i}, \epsilon^2}
                                         \normal\del{\Xmin \mid \min_i\cbr{X_i}, \epsilon^2}\,,
\end{split}
\label{eq:normal_lik}
\end{equation}
where \(\normal\del{x \mid \mu, \sigma^2}\) denotes the density of a normal distribution with mean \(\mu\) and variance \(\sigma^2\) evaluated at \(x\).
For small \(\epsilon\), this is a reasonable approximation enabling the use of MCMC techniques.

\begin{figure}[tbp]
\centering
\includegraphics[height=0.3\textheight,width=0.99\textwidth,keepaspectratio]{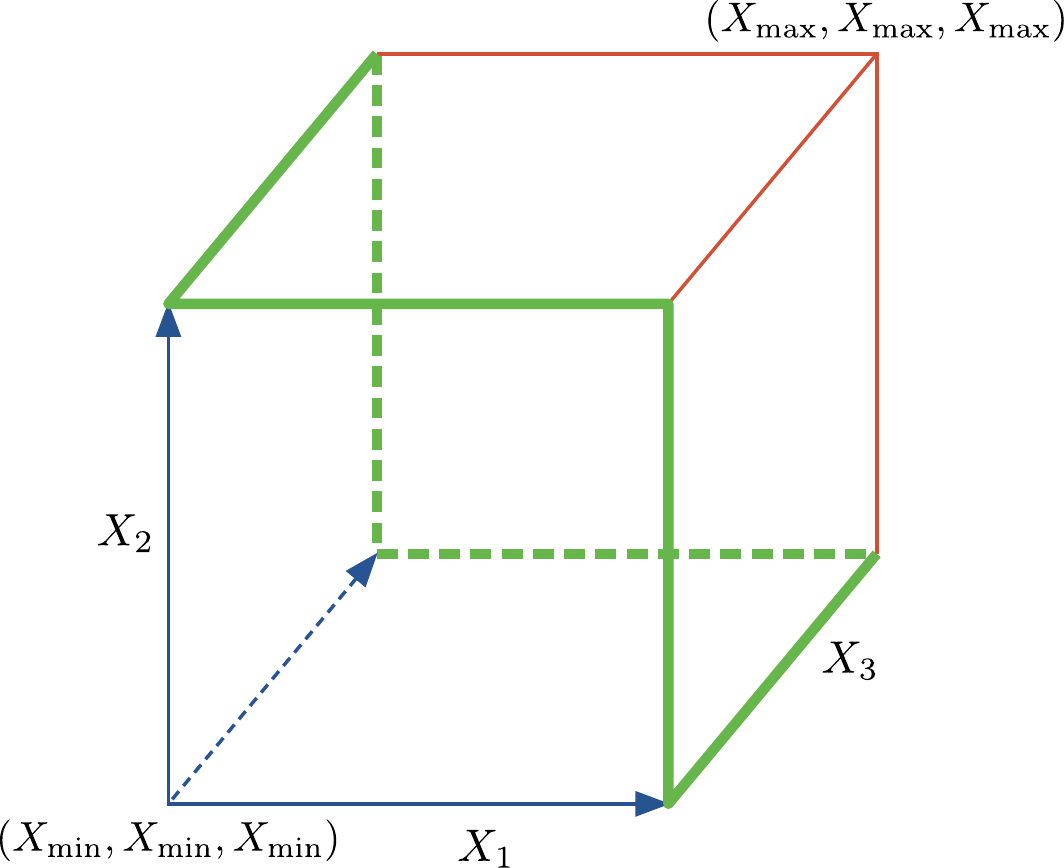}
\caption{\label{fig:constraints3d}
With three variables \(X_1\), and \(X_2\) and \(X_3\), \(\Fcond\) resides in the one-dimensional six-sided loop shown with thicker green lines. This is a 1D manifold embedded in 3D space, and possessing sharp corners, making it difficult for most MCMC algorithms to explore.}
\end{figure}

This approximation to \(\Fcond\) remains a difficult distribution to sample from.
We illustrate the constraint in a 3-dimensional setting in \autoref{fig:constraints3d}.
The MCMC must travel efficiently along the six edges of the allowed subspace,
and navigate corners when the index of the extremum components change.

Hamiltonian Monte Carlo (HMC) has shown a remarkable ability to navigate complicated distributions, including distributions where the typical set has ``pinch points'' of strong
curvature~\citep{betancourt2017conceptual}, similar to the ``corners`` in \(\Fcond\).
HMC's efficient sampling relies on gradient information in order to move towards regions of high probability.
The normal likelihood \autoref{eq:normal_lik} softened the extrema constraints,
but the maximum and minimum functions also remove information from the gradient.
The partial derivative of the log-likelihood of the maximum term with respect to \(X_i\) is proportional to:
\begin{equation}
\frac{\partial \log \normal\del{\Xmax \mid \max_i\cbr{X_i}, \epsilon^2}}{\partial X_i} \propto \del{\Xmax - X_i} \Ind\cbr{\argmax_j\del{X_j} = i} \,.
\end{equation}
The gradient pulls the maximum of the current MCMC state towards \(\Xmax\),
and ignores all other components.
This makes it difficult for HMC to efficiently explore scenarios where other components set the maximum.

In order to assist the HMC algorithm, we make another approximation.
We replace the \(\max\) and \(\min\) functions in \autoref{eq:normal_lik} with the \(\softmax\) and \(\softmin\) functions, defined on real inputs \(x_1, \dotsc, x_p\) as:
\begin{equation}
\begin{split}
    \softmax\del{x_1, \dotsc, x_p ; k} &= \frac{1}{k} \log\del{\sum_{i=1}^p e^{kx_i}}\,, \\
    \softmin\del{x_1, \dotsc, x_p ; k} &= -\softmax\del{-x_1, \dotsc, -x_p; k}\,.
\end{split}
\label{eq:softmax}
\end{equation}
As the sharpness parameter \(k\) goes to infinity, \(\softmax\) approaches the maximum, and \(\softmin\) approaches the minimum.
This substitution costs a small price in accuracy due to the approximation, but there is an important benefit: the gradient is now informative for all components of \(X\):
\begin{equation}
\frac{\partial \log \normal\del{\Xmax \mid \softmax\del{X_{1:p} ; k}, \epsilon^2}}{\partial X_i} \propto \del{\Xmax - \softmax\del{X_{1:p} ; k}} 
        \frac{e^{k X_i}}
             {\sum_{j=1}^p e^{k X_j}} \,.
\end{equation}
These modifications make HMC a viable algorithm to efficiently draw samples from the constrained posterior.
Setting \(k\) and \(\epsilon\) is a compromise between exactness and efficiency;
we found \(k=10\) and \(\epsilon=0.1\,\degreeC\) to perform well for our application.

Henceforth, we refer to this use of HMC and a smoothmax approximation to the target distribution as SmoothHMC.
SmoothHMC provides a generally applicable algorithm to draw from a multivariate distribution conditionally on the observed minimum and maximum of its components.

\subsection{Demonstration of SmoothHMC}
\label{sec:toy_example}

\begin{figure}[!tb]
\centering
\includegraphics[width=0.99\textwidth,height=0.55\textheight,keepaspectratio]{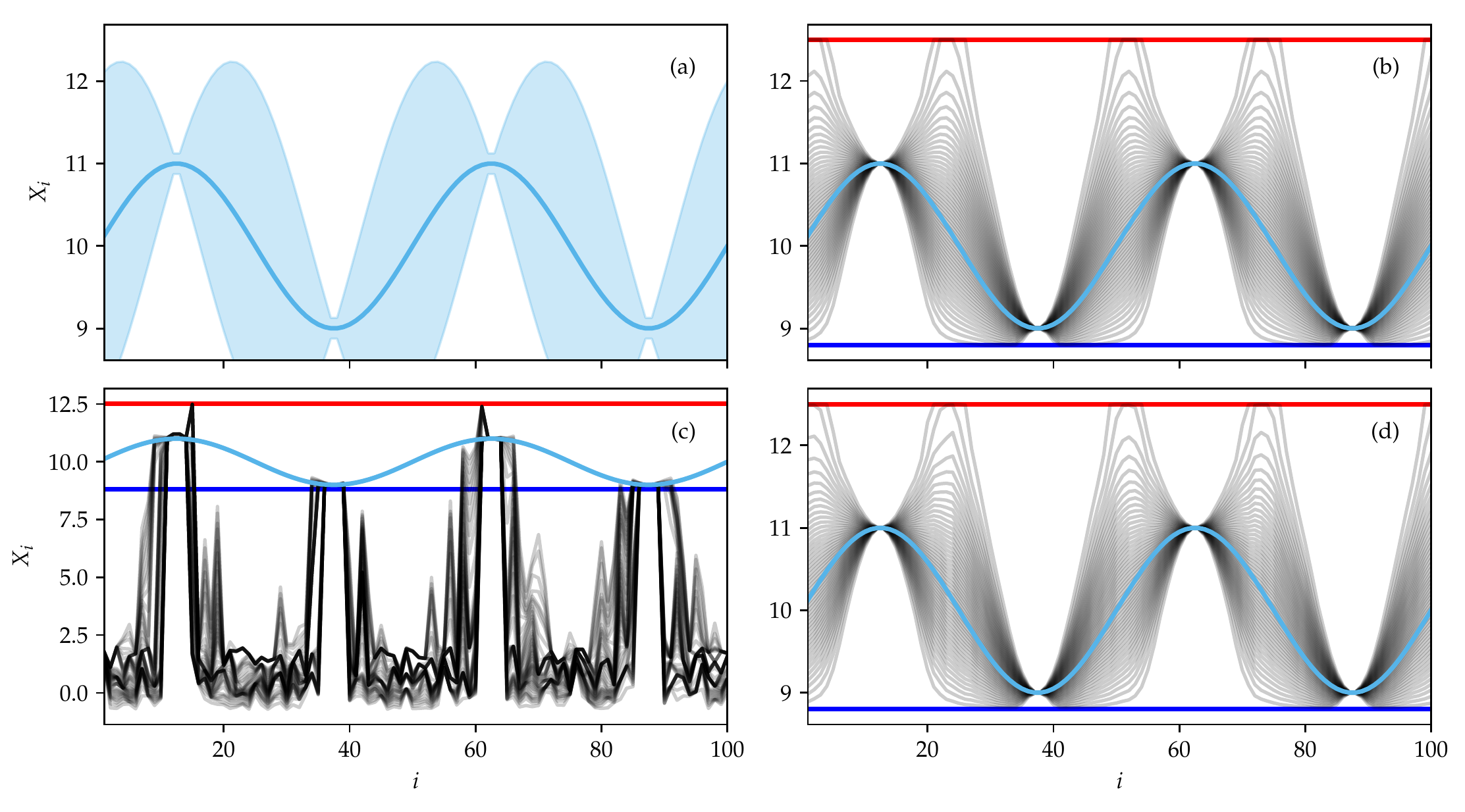}
\caption{\label{fig:toy_quantiles}(a) Prior distribution of \(X_i\) displayed as mean \(\mu_i\) (shown in every subplot to ease comparison) with 2~SD envelope; (b) Quantiles of the analytically derived posterior \(\Fcond\) conditioned on \(\Xmin\) (dark blue line) and \(\Xmax\) (red line); (c) Quantiles of the samples drawn from \(\Fcond\) using HMC (without the \(\softmax\) approximation); (d) Quantiles of the samples drawn from \(\Fcond\) using SmoothHMC.}
\end{figure}

We demonstrate SmoothHMC's ability to obtain draws from \(\Fcond\) in a simplified setting where the distribution function of \(\Fcond\) can be derived analytically and computed easily.
In our application, \(F_X\) is the posterior predictive multivariate normal distribution \(\T_\miss \mid \T_\obs\) obtained from nearby measurements, with mean and marginal variance evolving smoothly from one prediction to the next.
To parallel this, we specify a random vector \(X\) with each component \(X_i\) normally distributed, and with sinusoidal means and variances, but without any correlations between them,
so as to avoid a combinatorial explosion when obtaining \(\Fcond\) analytically:
\begin{equation}
\begin{split}
& X_i \overset{\indep}{\sim} \normal \del{\mu_i, \sigma_i} \,,\quad i = 1, \dotsc, 100 \,, \\
& \mu_i = 10 + \sin\del{2\pi i / 50} \,,\quad \sigma_i = 0.1+\cos^2\del{2\pi i / 50} \,.\\
& \Xmax = \max_i\cbr{X_i} \quad\text{and}\quad \Xmin = \min_i\cbr{X_i} 
\,.
\end{split}
\label{eq:toyspec}
\end{equation}
The unconstrained distribution of \(X_i\) is shown in \autoref{fig:toy_quantiles}(a).
In this example, we aim to sample from the distribution of \(X_i\) subject to the observation that \(\Xmax=12.5\) and \(\Xmin=8.8\).
An analytical derivation of the marginals of \(\Fcond\) is provided in \autoref{sec:analytical_posterior}, and its quantiles shown in \autoref{fig:toy_quantiles}(b).

To obtain samples from \(\Fcond\), we use the implementation of HMC provided by the Stan probabilistic programming language \citep{stancite}.
In Stan, the user specifies a probabilistic data-generating process for the observed data, based on parameters and latent variables with accompanying priors.
Stan then compiles this model into a custom \texttt{C++} program that efficiently implements
posterior sampling using HMC.
We implement two Stan models to draw from \(\Fcond\);
code for both is available from the GitHub account of the first author.
The first model implements \autoref{eq:normal_lik},
with a narrow normal likelihood term around the maximum and minimum,
while the second model also uses the \(\softmax\) approximation \autoref{eq:softmax}.
For each Stan model, we obtain 4 HMC chains each with 10,000 warm-up samples followed by 10,000 samples.
The quantiles of the samples obtained without the \(\softmax\) approximation are shown in \autoref{fig:toy_quantiles}(c).
By default, Stan initizializes each \(X_i\) uniformly at random between -2 and 2,
and for most variables, the algorithm remains stuck near the initial values.
Most samples do not conform to the constraints imposed by the observed \(\Xmin\) and \(\Xmax\) values, which invalidates these imputations.
However, once we replace the maximum function with the \(\softmax\) function,
with quantiles shown in \autoref{fig:toy_quantiles}(d),
SmoothHMC is able to draw samples that respect the observed extrema.
Furthermore, a visual comparison of the analytical quantiles in \autoref{fig:toy_quantiles}(a)
and the SmoothHMC sample quantiles in \autoref{fig:toy_quantiles}(d) confirms that
this sampling algorithm delivers a close approximation of the marginal distribution of each variable \(X_i\) in \(\Fcond\).
    
\begin{figure}[tbp]
\centering
\includegraphics[width=0.97\textwidth,height=0.7\textheight,keepaspectratio]{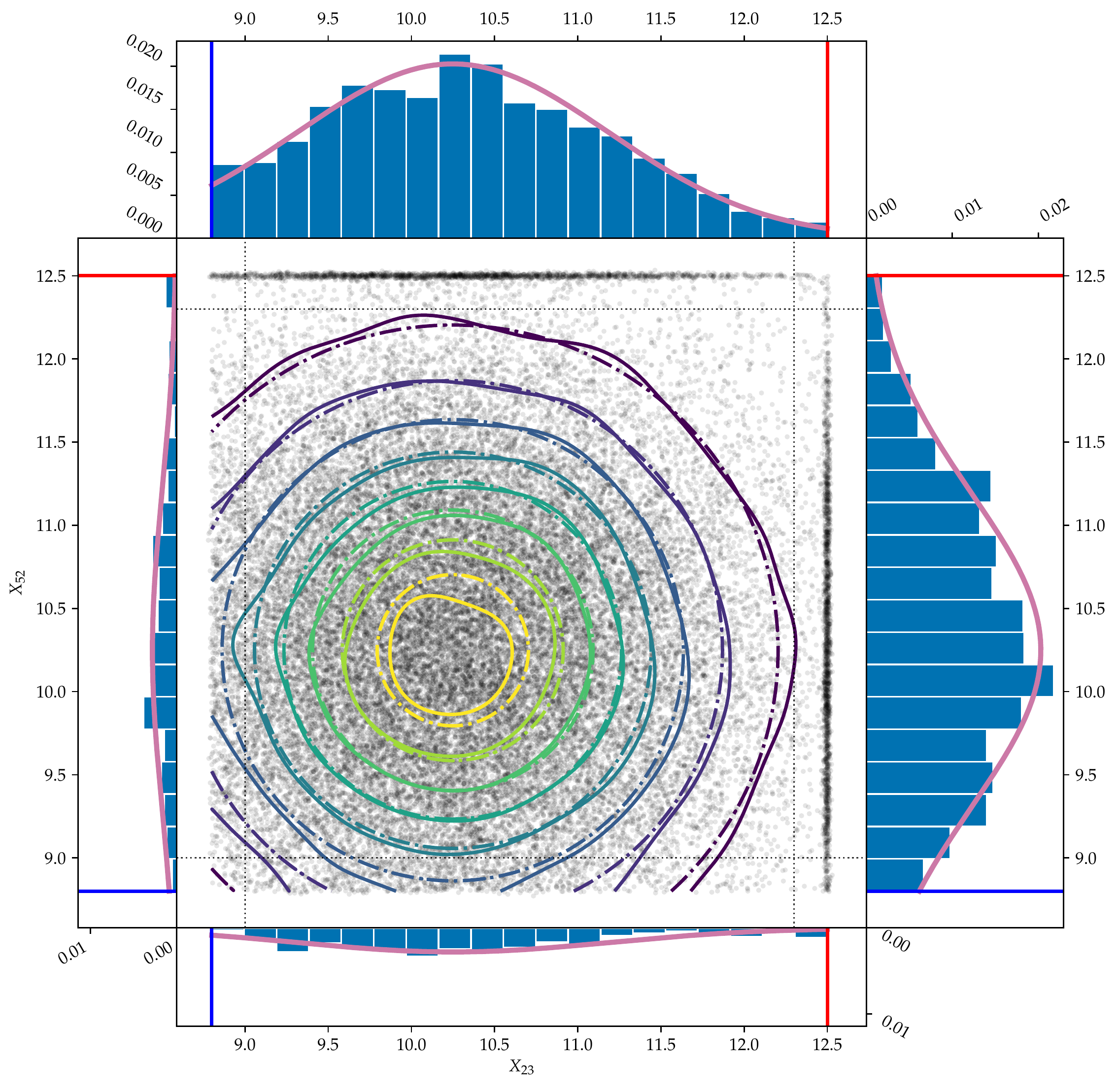}
\caption{\label{fig:toy_joint}
Comparison of the joint joint PDF of \(X_{23}\) and \(X_{52}\)
obtained analytically and from SmoothHMC samples.
The central scatterplot shows the 40,000 SmoothHMC samples.
Superimposed thereon are a contour plot (dash-dotted) of the joint marginal PDF of \(\Fcond\) for \(X_{23}\) and \(X_{52}\),
and a contour plot (solid lines) of kernel density estimates for the subset of SmoothHMC samples where neither \(X_{23}\) or \(X_{52}\) is the min or max,
obtained with a normal kernel with bandwidth \(0.2\)
(estimates are divided by the integrated mass of the kernel that is inside of the \(\Xmin\)/\(\Xmax\) boundaries).
The dotted lines are one bandwidth away from the \(\Xmin\)/\(\Xmax\) boundaries, beyond which kernel density estimates are less reliable.
The four histograms around the scatter plot are of the SmoothHMC samples adjacent to their x-axis, when one of the variables is an extremum.
For example, the top histogram is of \(X_{23}\) for samples where \(X_{52}\) is the max, 
while the super-imposed pink line is the (truncated normal) marginal PDF of \(X_{23}\) if it is neither the max nor the min,
times the probability that \(X_{52}\) is the max.
Blue and red lines indicate \(\Xmin\) and \(\Xmax\) respectively.
}
\end{figure}
    
We also visually verify that SmoothHMC samples correctly from the joint distribution of any combination of variables.
We do this for a pair of variables, \(X_{23}\) and \(X_{52}\), with results shown in \autoref{fig:toy_joint}.
There is a close match between the contours of the analytical joint distribution function (dash-dotted contour lines) and of the kernel density estimate (solid contour lines) of the SmoothHMC samples.
Each of the four histogram of samples where \(X_{23}\) or \(X_{52}\) occupies the minimum or maximum position matches the corresponding analytical distribution function well.
This visual comparison of the sample and analytical distributions shows that SmoothHMC is yielding a good approximation of a sample drawn from the true \(\Fcond\) in this example.
We did not examine the behavior of the sampling algorithm for the joint distribution of more than two variables due to the difficulty of visualizing such a distribution,
but we see no reason to suspect that the algorithm suffers from pathological behaviors that do not appear in these univariate and bivariate inspections.

\subsection{Smoothmax Temperature Model}

Armed with the SmoothHMC algorithm implemented in Stan, we now return to the problem of imputing hourly temperature measurements.
To impute the missing temperatures, we need to draw from the posterior distribution \(\T_\miss \mid \T_\obs, \Tn, \Tx\).
Bayes' theorem conditional on \(\T_\obs\) gives
\begin{equation}
    \Pr\del{\T_\miss \mid \T_\obs, \Tn, \Tx} = \frac{
        \Pr\del{\Tn, \Tx \mid \T_\obs, \T_\miss } 
        \Pr\del{\T_\miss \mid \T_\obs}
        }{
        \Pr\del{\Tn, \Tx \mid \T_\obs}
        }\,.
\end{equation}
The second term in the numerator is the posterior obtained in \autoref{sec:predict_nearby} now acting as a prior.
The denominator is a normalizing constant.
The first term in the numerator is either zero or one, indicating whether \(\T_\miss\) satisfies the constraint imposed by the observed \(\Tn\) and \(\Tx\).
Therefore, the posterior distribution takes a similar form to \autoref{eq:bayes_exact}, which motivates the use of SmoothHMC.

A small leap of faith is needed to accept that SmoothHMC's success in a toy example in \autoref{sec:toy_example} will extend to this application.
There are three important differences between the toy example and the temperature time series model.
Firstly, \(F_X\) is now a multivariate normal distribution with strong correlations obtained as the posterior distribution of a Gaussian process in \autoref{eq:unconstrained_post}.
Secondly, instead of a single minimum and maximum, we observe extrema for every 24 hour period.
Thirdly, we allow for the mean temperature to be different at different locations,
and so the imputed temperatures are shifted by an additional parameter \(\mu_{\miss}\),
to which we attach a vague prior.
To summarize, the probabilistic model that we wish to draw posterior imputations of \(\T_\miss\) from is given by:
\begin{equation}
\begin{split}
    \T_\miss &= \mu_\miss + \T_{\miss \mid \obs} 
        \quad\text{with}\quad 
        \mu_{\miss} \sim \normal\del{0,10^2}
    \,\text{, and} \\
    \T_{\miss \mid \obs} &= \T_{\miss} \mid \T_\obs \sim \normal\del{\mu_{\miss \mid \obs}, \Sigma_{\miss \mid \obs}}\, \\
    \del{\Tx}_{\iday} &= \max\cbr{\T_{\miss,\,i}\text{, for all \(i\) such that } t_{\miss\,,i} \in \dayset{\iday}} \,,\\
    \del{\Tn}_{\iday} &= \min\cbr{\T_{\miss,\,i}\text{, for all \(i\) such that } t_{\miss\,,i} \in \dayset{\iday}} \,.
\end{split}
\label{eq:idealmodel}
\end{equation}
To sample from this model with SmoothHMC, we modify it with the \(\softmax\) approximation to the maximum, and a normal likelihood:
\begin{equation}
\begin{split}
    \del{\Tx}_{\iday} &\sim \normal\del{\softmax_{i \in \dayset{\iday}} \cbr{ T_{\miss,i}; k=10}, 0.1^2}\,, \\
    \del{\Tn}_{\iday} &\sim \normal\del{\softmin_{i \in \dayset{\iday}} \cbr{ T_{\miss,i}; k=10}, 0.1^2}\,.
\end{split}
\label{eq:smoothed_model}
\end{equation}
A few samples from this imputation procedure are shown in \autoref{fig:imputations_2x2}(c).
From May 28, 2015 to June 1, 2015, hourly temperatures are imputed at KALO, using the hourly temperature measurements from nearby stations to inform the course of the temperatures, and constraining the imputations within the \(\Tx\) and \(\Tn\) extracted from the withheld time series at 11:00 each day.
Imputations are obtained in nine day windows for computational reasons, with three days of overlap between adjacent windows so each imputation can be made at least three days away from the window's edges.
One can verify visually that the imputations respect the \(\Tn\) and \(\Tx\) constraints, reaching but not exceeding each extreme on each day.
Since we actually have hourly data for KALO, yet fed our algorithm only the daily extremes, we can also plot the hidden temperatures (in black), and see how faithfully the imputations reproduce them.
We see that the imputations indeed track the true measurements closely.
This success demonstrates that SmoothHMC is capable of imputing temperature time series from the constrained posterior distribution \(\T_\miss \mid \T_\obs, \Tn, \Tx\).

\section{Model Diagnostics}
\label{sec:diagnostics}

\subsection{Variogram}\label{variogram}

Model fit can be visually inspected by plotting temporal and spatial semi-variograms. 
The semi-variogram  of a stationary spatio-temporal function \(Y(\xvec,t)\) is a function of the spatial lag \(\hvec\) and the temporal lag \(r\) \citep[see for example][chapter 6]{sherman2011spatial}:
\begin{equation}
    \gamma\del{\hvec,r} = \frac{1}{2} \E\sbr{\del{Y\del{\xvec,t}-Y\del{\xvec+\hvec,t+r}}^2}
\end{equation}
For a Gaussian Process model, with a stationary covariance function \(k(\hvec,r)=k(\xvec,\xvec+\hvec,t,t+r)\), this can be expressed as:
\begin{equation}
    \label{eq:gp_variogram}
    \gamma\del{\hvec,r} = \sigman^2 + k\del{0,0} - k(\hvec,r)\,.
\end{equation}
From the data, the variogram can also be estimated empirically, by averaging the square differences of any two observations that are separated by \(\hvec\) in space, and \(r\) in time (in practice, time lags are binned).
By comparing the empirical variogram to the variogram of the fitted covariance, 
we obtain a visual diagnosis of the model.

\begin{figure}[!tbp]
\centering
\includegraphics[width=0.99\textwidth]{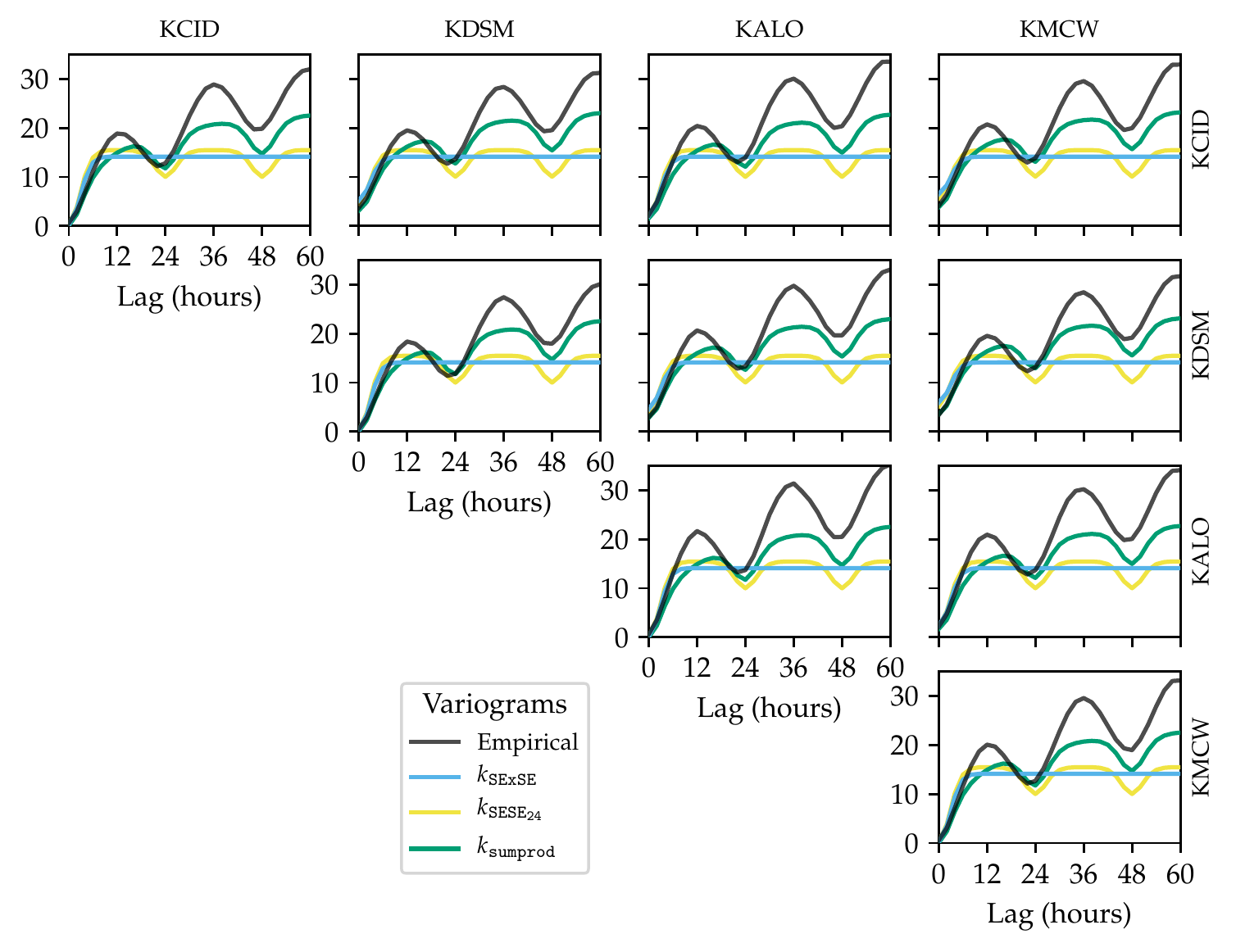}
\caption{\label{fig:spatial_variogram}
Semi-variograms of the temperature temperature time series at four Iowa weather stations, each labeled by its ICAO code.
The empirical semi-variograms are shown in black, and the fitted variograms for the three covariance models proposed in this paper are shown in color.
The temporal semi-variograms are shown on the diagonal, while the off-diagonal plots show the semi-variograms as a function of time lag for a fixed distance \(\hvec\) equal to the distance between the two stations.
}
\end{figure}

In our Iowa example, there are only four possible locations. For each location, we plot the empirical temporal variogram \(\widehat\gamma\del{0,r}\). Then, for each pair of stations separated by \(\hvec\) (fixed), we can also plot the estimate \(\widehat\gamma\del{\hvec,r}\). 
We overlay the model's semi-variogram from equation \autoref{eq:gp_variogram}, resulting in \autoref{fig:spatial_variogram}.
For each variogram, we have removed the effect of the \(k_\mu\) covariance, which would shift the variogram between two stations by a large arbitrary constant.
Correspondingly, we subtract the mean of each observed time series before obtaining the empirical variogram.

We notice that the variogram of the model with product covariance \autoref{eq:ksese} tracks the empirical variogram well at short lags, but fails to capture the periodicity in the empirical variogram, and the fit degrades at long lag. We improve the model in \autoref{sec:improving_model}.

\subsection{Error and Expected Error}

The variogram gives us a visual diagnostic of the overall model fit. 
To quantify the model's predictive ability in the Iowa example, we compare the posterior mean temperature to the withheld truth, and obtain the empirical mean squared error (MSE) for $N$ predictions as:
\begin{equation}
    \label{eq:mse}
    \frac{1}{N} \sum_{i=1}^N \sbr{
        \T_{\miss,i}
        -
        \E\del{\T_{\miss,i} \mid \T_\obs,\Tx,\Tn}
}^2\,.
\end{equation}
Equation~\autoref{eq:mse} is for the final predictions obtained using nearby hourly temperatures and local daily maxima and minima.
A similar diagnostic can be computed for the intermediary predictions, which exclude the local \(\Tx\) and \(\Tn\) information.
At that stage, we are not concerned with any overall bias in the predicted temperatures, so we instead compute the sample variance of the errors as
\begin{equation}
    \var\del{\error} = \var_i \cbr{
        \T_{\miss,i}
        -
        \E\del{\T_{\miss,i} \mid \T_\obs}
    }\,.
    \label{eq:varerr}
\end{equation}

\begin{table}[tbp]
\begin{center}
\bgroup
\def\arraystretch{1.1}
\begin{tabular}{lrrrrr}
\hline
Model & \multicolumn{1}{l}{log-} & \(\var(\error)\) & \(\widehat{\var}(\error)\) & \(\mse\) & \(\widehat{\mse}\) \\
& \multicolumn{1}{l}{likelihood} & \multicolumn{1}{c}{\autoref{eq:varerr}} & \multicolumn{1}{c}{\autoref{eq:expected_varerr}} & \multicolumn{1}{c}{\autoref{eq:mse}} & \multicolumn{1}{c}{\autoref{eq:expected_mse}} \\
\hline
\(\kSESE\) & -55,614 & 1.59 & 0.88 & 1.12 & 0.44\\
\(\kdiurn\) & -54,472 & 1.63 & 0.97 & 1.12 & 0.69\\
\(\ksumprod\) & -45,944 & 1.32 & 1.19 & 1.04 & 0.81\\
\hline
\end{tabular}
\caption{
    Model diagnostics for three Gaussian process covariance functions. 
    \label{table:diagnostics}
}
\egroup
\end{center}
\end{table}

For our purposes, it isn't sufficient for the spatio-temporal model to yield good predictions; we also require a good estimate of its own accuracy.
We estimate the error variance expected by the model by sampling random draws \(\T^{(k)}_\miss\), \(k=1,\dotsc,K\) from the multivariate normal posterior distribution \(\T_{\miss,i} \mid \T_\obs\), and computing the variance between the samples and the posterior expectation:
\begin{equation}
\widehat{\var}\del{\error} = \frac{1}{K} \sum_{k=1}^K \var_i \cbr{\T^{(k)}_{\miss,i} - \E\del{\T_{\miss,i} \mid \T_\obs}}
\label{eq:expected_varerr}
\end{equation}
Similarly, to estimate the MSE expected by the model we use the MCMC draws \(\tilde\T^{(k)}_\miss\), \(k=1,\dotsc,K\) from SmoothHMC, and compute the MSE between the samples and the posterior expectation:
\begin{equation}
    \widehat{\mse} = \frac{1}{K} \sum_{k=1}^K \frac{1}{N} \sum_{i=1}^N \sbr{\tilde\T^{(k)}_{\miss,i} - \E\del{\T_{\miss,i} \mid \T_\obs,\Tx,\Tn}}^2
\label{eq:expected_mse}
\end{equation}
When evaluating models, we want the errors to be small, and so the error variance and MSE to be low. 
A well-calibrated model should also have the estimated error variance \autoref{eq:expected_varerr} and MSE \autoref{eq:expected_mse} close to their empirical values \autoref{eq:varerr} and \autoref{eq:mse} respectively.

These diagnostics for our first spatio-temporal model, the product of squared exponentials, are found in the first row of \autoref{table:diagnostics}. The error variance using only nearby measurements is already fairly low, with typical errors of order \(\sqrt{1.59}=1.26\,\degreeC\). Incorporating \(\Tn\) and \(\Tx\) using SmoothHMC reduces it further to \(\sqrt{1.12}=1.06\,\degreeC\). 
However, the model is overly optimistic, and the expected errors underestimate the empirical errors.

\section{Improving the Basic Model}
\label{sec:improving_model}

In this section, we develop more sophisticated Gaussian process covariances than the simple product of squared exponential kernels \(\kSESE\) \autoref{eq:ksese}. 
We then assess whether these models improve the variogram and the predictive diagnostics that we presented in \autoref{sec:diagnostics}.

The most salient feature of the empirical variogram that is not captured by the \(\kSESE\) covariance is the oscillation with a 24-hour period.
It is intuitively clear that the diurnal cycle induces this periodic covariance, and that our model should be improved by incorporating this feature.
Gaussian processes allow for periodic components of the covariance, for example the periodic squared exponential covariance function, which we use with a 24-hour period
\begin{equation}
    k_{24}(t,t') = \sigma_{24}^2 \exp\sbr{ - \frac{2}{\ell_{24}^2} \sin^2\del{
        \pi \frac{t-t'}{\text{24 hrs}} 
        }}\,.
\end{equation}
We modify the spatiotemporal model by adding this diurnal component to it, with its own spatial decay component \(k_{\subspace{}24}\) (with the same form as \(k_{\subspace}\) in \autoref{eq:kspace}, and again with variance parameter fixed to 1):
\begin{equation}
    \kdiurn(\xvec,\xvec',t,t') = k_{\subtime}(t,t') \cdot k_{\subspace}(\xvec, \xvec') 
        + k_{24}(t,t') \cdot k_{\subspace{}24}(\xvec, \xvec')
        + k_\mu(\xvec, \xvec') 
        \,.
\end{equation}
We also propose a more complex model, which breaks up \(k_{\subtime}\) into short-term, medium-term and long-term correlation components:
\begin{equation}
\begin{aligned}
    \ksumprod(\xvec,\xvec',t,t') &= 
           k_{\subtime{}1}(t,t') \cdot k_{\subspace{}1}(\xvec, \xvec')  &\text{(short-term variation)} \\
        &+ k_{\subtime{}2}(t,t') \cdot k_{\subspace{}2}(\xvec, \xvec')  &\text{(medium-term variation)} \\
        &+ k_{\subtime{}3}(t,t') \cdot k_{\subspace{}3}(\xvec, \xvec')  &\text{(long-term variation)} \\
        &+ k_{24}(t,t') \cdot k_{\subspace{}24}(\xvec, \xvec') &\text{(diurnal cycle)} \\
        &+ k_\mu(\xvec, \xvec') &\text{(station mean)}
\end{aligned}
\label{eq:sumprod_kernel}
\end{equation}
Each of \(k_{\subtime{}1}\), \(k_{\subtime{}2}\), and \(k_{\subtime{}3}\), is a rational quadratic kernel:
\begin{equation}
    k_{RQ}(t,t') = \sigma^2 \del{1 + \frac{\del{t-t'}^2}{2\alpha\ell^2} }^{-\alpha}
\end{equation}
and is multiplied by a spatial decay component, specified as a squared exponential \autoref{eq:kspace} with variance fixed at 1.
Fitted covariance parameters for \(\kdiurn\) and \(\ksumprod\) are found in \autoref{table:fitted_params}.

We now have three competing Gaussian process models, with covariance functions \(\kSESE\), \(\kdiurn\), and \(\ksumprod\) respectively. 
We can compare them in four ways.
Firstly, the variogram fit in \autoref{fig:spatial_variogram} is visibly improved by the introduction of the the diurnal component in \(\kdiurn\), and by the additional spatio-temporal correlation decay components in \(\ksumprod\).
Secondly, the marginal log-likelihood is the quantity maximized by the parameter fitting procedure in \autoref{eq:optimization}, with maximized values found in the second column of \autoref{table:diagnostics}.
The more complex models indeed yield a higher log-likelihood, promising a better model fit which should yield better predictions.
Thirdly, we compare the variance of the error in the predicted temperatures specified in \autoref{eq:varerr} when withholding all the data from a test station. 
Averaged over all of 2015, this is given in the third column of \autoref{table:diagnostics}, and shows more mixed results.
The diurnal model \(\kdiurn\) performs slightly worse than the simple \(\kSESE\) model, and \(\ksumprod\) only yields a small improvement.
Fourthly, we compare the mean squared error specified in \autoref{eq:mse} for imputations at the test station incorporating \(\Tn\)/\(\Tx\). 
Results in the fifth column also show more modest improvements for the more complex models.
That said, with an expected MSE closer to its true value, \(\ksumprod\) does give better estimates of its own inaccuracy.

We interpret these results as a reminder that prediction accuracy using Gaussian process is sensitive to model specification when extrapolating, but fairly insensitive to the model when interpolating \citep{stein2012interpolation}.
Our imputations interpolate the temperatures from nearby stations, further aided by the constraints imposed by the daily \(\Tn\) and \(\Tx\) measurements, which could explain why the choice of model does not seem to have a large impact on the performance of our imputation procedure.
This insensitivity can be seen as reassuring, as it shows robustness against model misspecification.

\section{Imputed Summary Statistics}

\autoref{fig:imputations_2x2}(d) shows the imputations produced under the \(\ksumprod\) covariance \autoref{eq:sumprod_kernel}.
This is the primary output of our imputation method, and the results are promising.
Firstly, just like in the toy example presented in \autoref{sec:toy_example}, the individual imputations meet the three constraints imposed by the measured minimum and maximum.
Each day, the imputations stay between \(\Tn\) and \(\Tx\),
and the temperatures always drop to \(\Tn\) and rise to \(\Tx\) at some time of the day.
The imputations reflect the uncertainty in the time at which the extrema are reached.
Notably, on some days, the posterior distribution of the warmest (or coldest) time is bimodal:
during the May 31 measurement window (from May 30 at 11:00 to May 31 at 11:00) for example, 72.5\% of SmoothHMC imputations reach their peak before 20:00 on May 30, 27.5\% after 8:00 on May 31, and none in between. 
We view as a particular strength of our approach that the imputations are able to capture this ambiguity, rather than being restricted to a single mode of the posterior distribution.

\begin{figure}[tbp]
\centering
\includegraphics[height=0.35\textheight,width=0.99\textwidth,keepaspectratio]{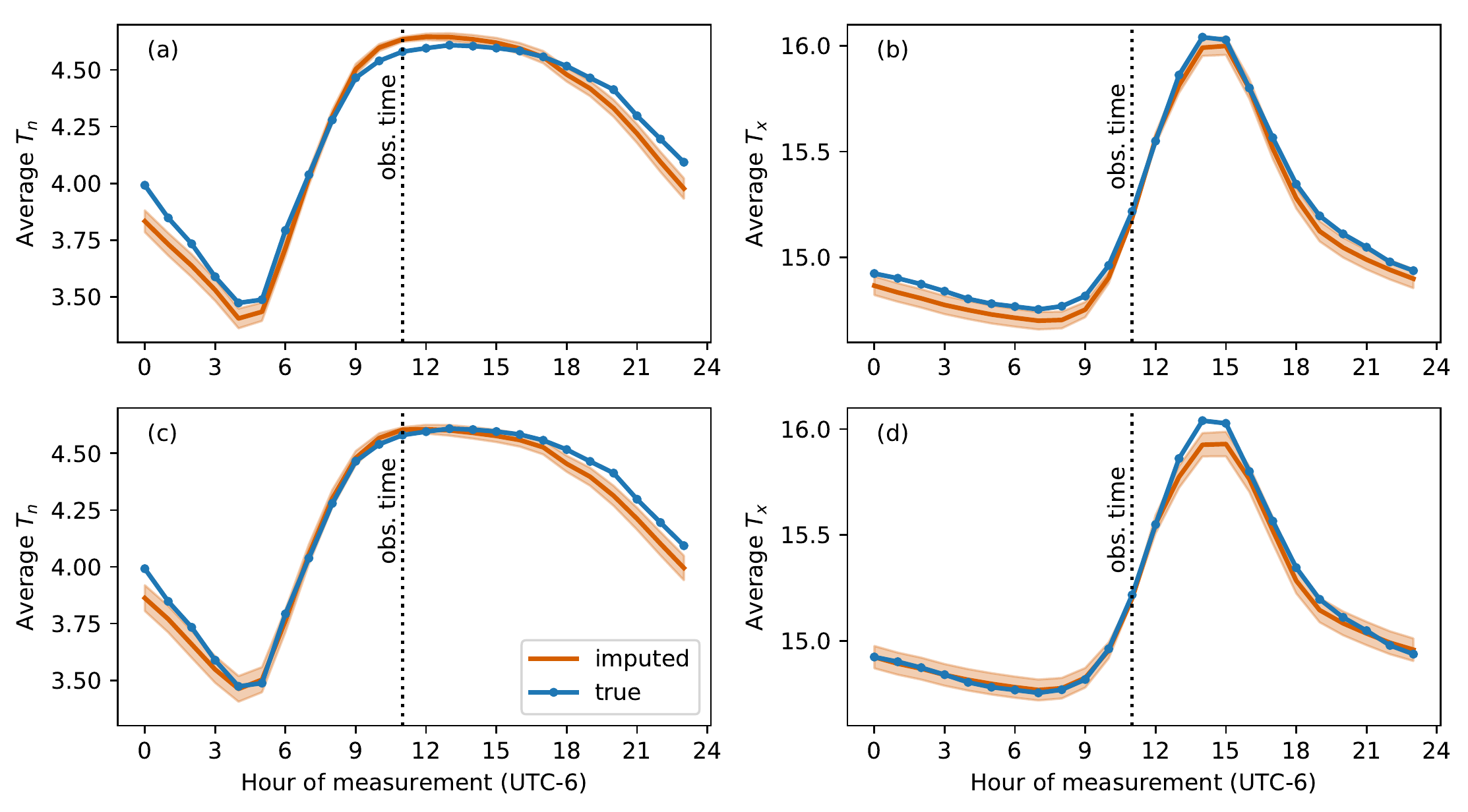}
\caption{\label{fig:imputed_summary_stats}
Average minimum (left) and maximum (right) daily temperature obtained under varying hour of measurement from KALO data (shown in blue), and from imputations of the withheld data (shown in orange with 2~SD envelope) obtained under the \(\kSESE\) covariance function (top) and the \(\ksumprod\) covariance function (bottom).}
\end{figure}
    
These imputations however are not the final aim of our analysis.
Rather, our stated goal is to undo, or at least account for, the sensitivity of summary statistics to measurement time, for example the average \(\Tx\) in \autoref{fig:waterloo_avgTnTx}.
Equipped with these imputations, is it possible to infer what the value of the summary statistic would have been for different measurement hours?
This possibility is demonstrated in \autoref{fig:imputed_summary_stats},
which shows the same summary statistic as in \autoref{fig:waterloo_avgTnTx} applied to the imputations as well as the (withheld) hourly data at KALO.
It can be seen that the imputed summary statistics track within about 0.1~\(\degreeC\) of the true values.
The product covariance \(\kSESE\) and the sum of products covariance \(\ksumprod\) seem to perform equally well imputing the summary statistics for different times, but the \(\ksumprod\) gives more honest, wider credible intervals.

\section{Inference on Measurement Hour}\label{inference-on-measurement-hour}

\begin{figure}[tbp]
\centering
\includegraphics[width=0.9\textwidth,height=0.4\textheight,keepaspectratio]{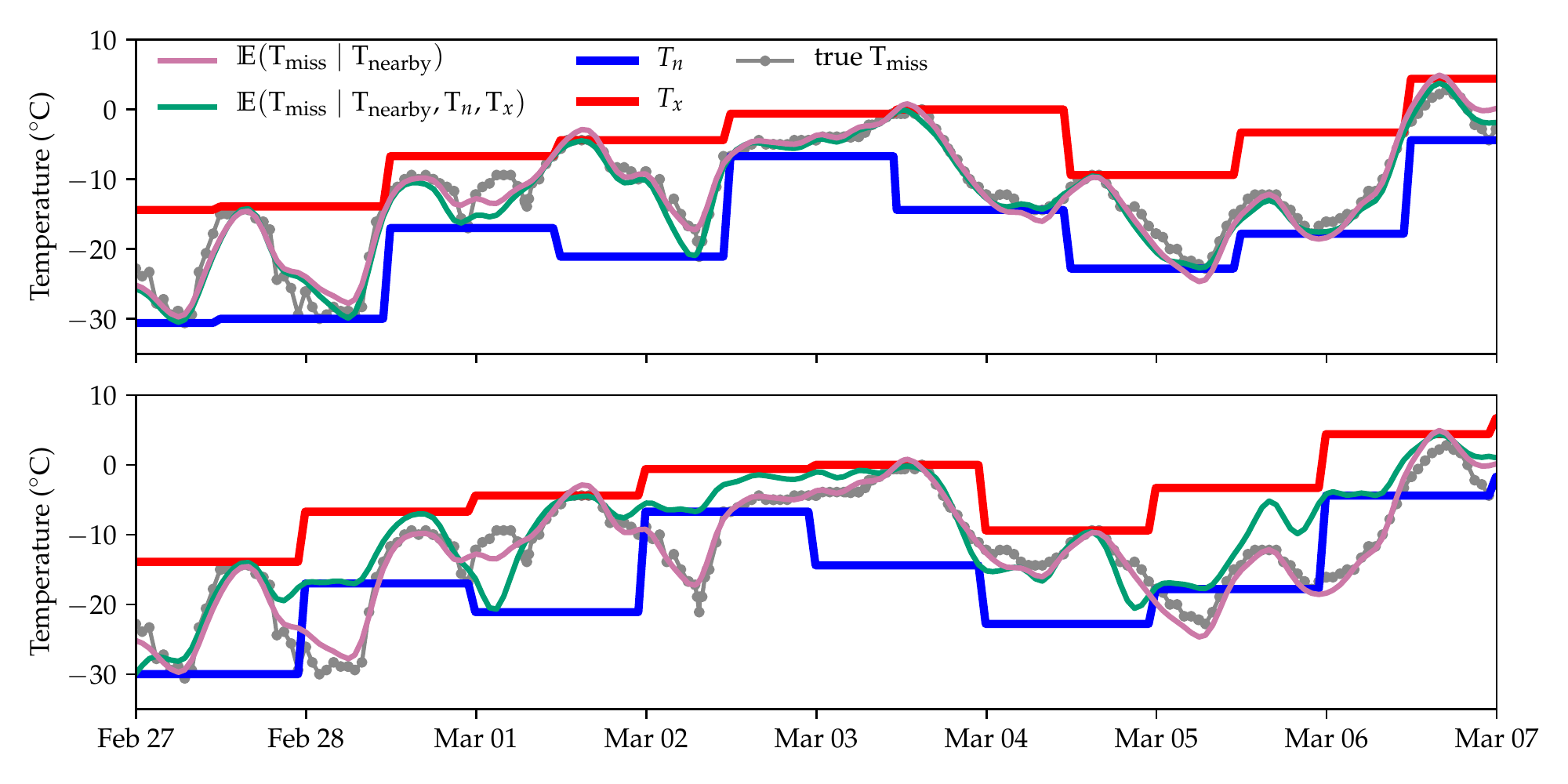}
\caption{\label{fig:measure_hour_example} Constrained and unconstrained imputations in an eight-day window, assuming (top) the correct measurement hour (11:00 UTC-6), and (bottom) a wrong measurement hour (23:00 UTC-6). Assuming the wrong measurement time drives the constrained mean imputation away from the unconstrained mean imputation.}
\end{figure}

Our analysis thus far has focused on the case where the hour of measurement \(\hour\) is known in advance.
This is a sometimes unrealistic assumption, and so inference on \(\hour\) is desirable.
It is conceptually straightforward to modify the measurement model \autoref{eq:smoothed_model} with a uniform prior on \(\hour\).
However, \(\hour\) affects which observations are attributed to each day's measurements, which has a 
discontinuous (observations suddenly jump from one day to the next) and non-differentiable effect on the posterior, 
and so Hamiltonian Monte Carlo becomes unviable.
We therefore do not consider the introduction of a uniform prior on \(\hour\) in Stan to be feasible.

\begin{figure}[tbp]
\centering
\includegraphics[height=0.4\textheight,width=0.9\textwidth,keepaspectratio]{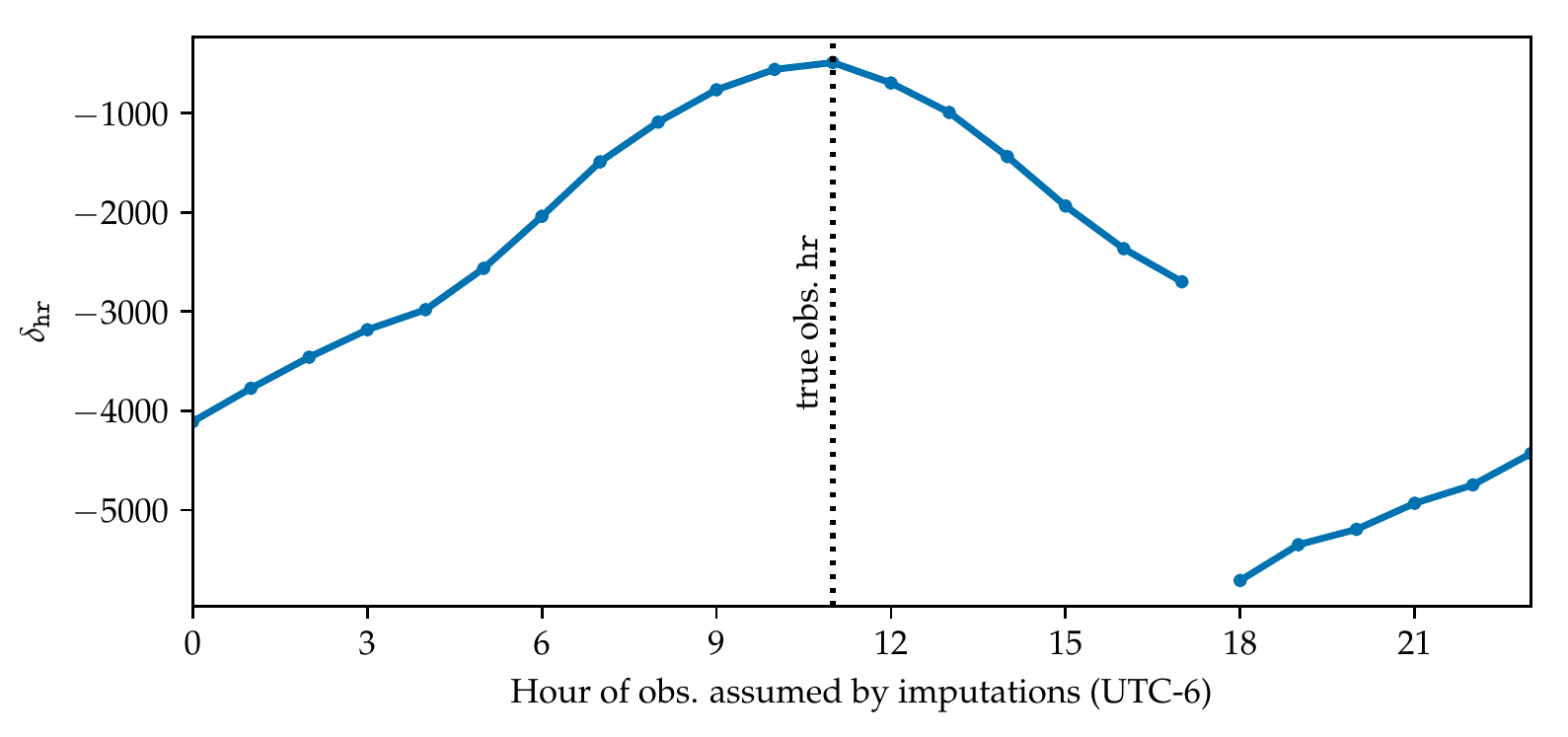}
\caption{
    \label{fig:hr_inference}
    Concordance \(\concordance_\hour\) for imputations of temperatures at KALO assuming measurement hours \(\hour=1,\dotsc,24\). 
    The true hour of measurement is 11:00, and obtains the highest \(\delta_\hour\).
    Observations are associated with the date on which the observation occurs in the UTC timezone, which causes the discontinuity at 18:00 UTC-6.
    }
\end{figure}

Our procedure allows us to obtain imputation samples of \(\T_\miss\) conditional on \(\T_\obs\), \(\Tn\), \(\Tx\) and \(\hour\).
If we do so for \(\hour=0,\dotsc,23\), is there information available in these samples to infer \(\hour\)?
We examine sample imputations in \autoref{fig:measure_hour_example} to gain intuition.
Rather unsurprisingly, assuming an incorrect measurement time leads to wildly inaccurate imputations, for example on March 2nd.
But notice also that assuming the wrong time causes the mean constrained imputation to depart further from the unconstrained imputation
(that is, the green and orange lines are further apart).
This can be interpreted as an indication of an incompatibility between \(\T_\obs\) and \(\Tn\)/\(\Tx\), caused by assuming the wrong \(\hour\).
We therefore propose to calculate the probability \(\concordance_\hour\) of the mean constrained imputation under the unconstrained posterior given by \(\autoref{eq:unconstrained_post}\), which we interpret as a measure of concordance between \(\T_\obs\) and \(\Tn\)/\(\Tx\):
\begin{equation}
    \concordance_\hour = \log \Pr\del{ \T_\miss \! = \! \mu\del{\hour} \mid \T_\obs }
    \,\text{, where}\ \,
    \mu\del{\hour} = \E\del{ \T_\miss \mid \T_\obs, \Tn, \Tx, \hour }\,,
\label{eq:concordance}
\end{equation}
Intuitively, \(\concordance_\hour\) will drop when the wrong \(\hour\) is assumed,
and we may be able to infer the true \(\hour\) by maximizing \(\concordance_\hour\).
In \autoref{fig:hr_inference}, we demonstrate this method on the withheld KALO time series, which
has been replaced by \(\Tx\)/\(\Tn\) observations made at \(\hour=11\).
We use SmoothHMC to impute the withheld data for all of 2015 under each possible measurement hour \(\hour=0,1,\cdots,23\).
For each set of imputations, we compute the posterior mean \(\mu\del{\hour}\) from the SmoothHMC samples, 
and the concordance \(\concordance_\hour\) \autoref{eq:concordance} (by necessity, modified to treat the center of each 73-day prediction window as an independent prediction).
Pleasantly, the concordance is highest when the true hour of measurement is used so that, in this example at least, the correct hour of measurement would be infered.

\section{Conclusion}\label{sec:conclusion}

Climatological research relies on the ability to track small changes over long periods.
For this reason, the bias induced by the measurement time
that we demonstrate in \autoref{sec:illustrate_bias}
could lead to wrong estimates and conclusions regarding long-term trends in temperature records.
We reformulated the source of this bias as a missing data problem, and imputed the missing hourly temperatures at the weather station using posterior samples from a spatiotemporal Gaussian process model.
The model allows the combination of information from the measured daily minimum (\(\Tn\)) and maximum (\(\Tx\)) temperatures, and from measurements of hourly temperatures at nearby meteorological stations.
While ours is not a physical model, it is very flexible, and it performs well for the task of interpolating temperatures between nearby locations and times.
Indeed, more complex covariance functions (with a diurnal component and a sum of short-range and long-range components) showed only modest improvements in the mean squared error of the imputations compared to a withheld hourly temperature record.

Our model accounts for miscalibration and bias in the hourly temperature measurements by assigning a mean parameter to each location, which is given a weak independent prior with no spatial correlation.
Therefore, our model only makes predictions at new locations up to a constant shift, and it only extracts information about the trajectory of the temperature time series from each weather station.
However, our strategy rests on the assumption that the trajectory is not affected by biases and miscalibration.
This assumption is violated for example if the presence of an airport has a very different effect on measured temperatures during the day and during the night, which would introduce bias in the imputations.
Our model could be improved in the future with a more complete characterization of how daily temperatures differ systematically between locations.

In order to condition the imputations on the daily \(\Tx\) and \(\Tn\), we developed SmoothHMC, a general algorithm based on Hamiltonian Monte Carlo with a smoothed approximation of the target distribution that can sample from a multivariate distribution conditionally on its observed minimum and maximum.
It showed an excellent ability to sample from the conditional distribution in an example where the distribution function can also be obtained analytically.
SmoothHMC is the main technical contribution of this paper, and we believe the method could find applications beyond the present setting.

We used this method to obtained imputations of the temperature time series that satisfied the constraints imposed by the measured \(\Tn\) and \(\Tx\).
The imputation of withheld temperatures at KALO track the true temperatures, within a root mean square error of \(1.02\,\degreeC\).
We view as particularly encouraging that the imputations successfully capture the uncertainty and sometimes bimodality in the time of the maximum or minimum temperature on days where this time is difficult to infer from the available information.

Future improvements to the imputation strategy would include the inclusion of rounding errors in the measurement model, explicit treatment of non-stationarity due to coastlines or other geographical features, and of altitude differences.
Gaussian process modeling allows for much flexibility in the choice of covariance kernels, and improved modeling should lead to more accurate imputations.

The imputed time series are the primary output of this work, but they are intended as a starting point for further analyses motivated by different scientific goals.
In particular, summary statistics can be applied to the imputations, such as the average \(\Tx\), under different choices of daily measurement hours.
Using imputations obtained for the withheld time series at KALO, we have demonstrated a good ability to recover this information (\autoref{fig:imputed_summary_stats}).
The average \(\Tx\) or \(\Tn\) is an example of a possible follow-up analysis, chosen mostly as an illustrative proof of concept.
We plan to use this method to compare the average temperature to the average of the measured \(\Tn\) and \(\Tx\) for a given location and year, with the former estimated using imputed time series.

Lastly, we discussed the possibility of inferring the hour of measurement \(\hour\).
We gave some intuition for maximizing the concordance \autoref{eq:concordance} in order to infer \(\hour\), and a single example where this strategy is successful.
While promising, we lack a theoretical justification for this approach.
It remains to be seen whether our approach is generalizable and successful in other examples, and whether it can be placed on sound theoretical bases.
Ideally we would wish to estimate the posterior probability \(\Pr\del{\hour \mid \T_\obs,\Tn,\Tx}\), for example by sampling from the joint posterior of \(\T_\miss, \hour \mid \T_\obs,\)\(\Tn\),\(\Tx\), but this is computationally difficult.
Furthermore, it would be desirable not merely to infer the hour of measurement for an entire year, but to detect changepoints: days on which the measurement practice changed from one hour of measurement to another.
We leave these improvements to inference of the measurement hour to future work.

\bibliographystyle{chicago}
\bibliography{temper}

\begin{thebibliography}{}

\bibitem[\protect\citeauthoryear{Baker}{Baker}{1975}]{baker1975effect}
Baker, D.~G. (1975).
\newblock Effect of observation time on mean temperature estimation.
\newblock {\em Journal of Applied Meteorology\/}~{\em 14\/}(4), 471--476.

\bibitem[\protect\citeauthoryear{Banerjee, Gelfand, Finley, and Sang}{Banerjee
  et~al.}{2008}]{banerjee2008gaussian}
Banerjee, S., A.~E. Gelfand, A.~O. Finley, and H.~Sang (2008).
\newblock Gaussian predictive process models for large spatial data sets.
\newblock {\em Journal of the Royal Statistical Society: Series B (Statistical
  Methodology)\/}~{\em 70\/}(4), 825--848.

\bibitem[\protect\citeauthoryear{Betancourt}{Betancourt}{2017}]{betancourt2017conceptual}
Betancourt, M. (2017).
\newblock A conceptual introduction to hamiltonian monte carlo.
\newblock {\em arXiv preprint arXiv:1701.02434\/}.

\bibitem[\protect\citeauthoryear{Carpenter, Gelman, Hoffman, Lee, Goodrich,
  Betancourt, Brubaker, Guo, Li, and Riddell}{Carpenter
  et~al.}{2017}]{stancite}
Carpenter, B., A.~Gelman, M.~Hoffman, D.~Lee, B.~Goodrich, M.~Betancourt,
  M.~Brubaker, J.~Guo, P.~Li, and A.~Riddell (2017).
\newblock Stan: A probabilistic programming language.
\newblock {\em Journal of Statistical Software, Articles\/}~{\em 76\/}(1),
  1--32.

\bibitem[\protect\citeauthoryear{Della-Marta and Wanner}{Della-Marta and
  Wanner}{2006}]{della2006method}
Della-Marta, P. and H.~Wanner (2006).
\newblock A method of homogenizing the extremes and mean of daily temperature
  measurements.
\newblock {\em Journal of Climate\/}~{\em 19\/}(17), 4179--4197.

\bibitem[\protect\citeauthoryear{Ducr{\'e}-Robitaille, Vincent, and
  Boulet}{Ducr{\'e}-Robitaille et~al.}{2003}]{ducre2003comparison}
Ducr{\'e}-Robitaille, J.-F., L.~A. Vincent, and G.~Boulet (2003).
\newblock Comparison of techniques for detection of discontinuities in
  temperature series.
\newblock {\em International Journal of Climatology\/}~{\em 23\/}(9),
  1087--1101.

\bibitem[\protect\citeauthoryear{Easterling, Peterson, and Karl}{Easterling
  et~al.}{1996}]{easterling1996development}
Easterling, D.~R., T.~C. Peterson, and T.~R. Karl (1996).
\newblock On the development and use of homogenized climate datasets.
\newblock {\em Journal of climate\/}~{\em 9\/}(6), 1429--1434.

\bibitem[\protect\citeauthoryear{Karl, Williams~Jr, Young, and Wendland}{Karl
  et~al.}{1986}]{karl1986model}
Karl, T.~R., C.~N. Williams~Jr, P.~J. Young, and W.~M. Wendland (1986).
\newblock {A model to estimate the time of observation bias associated with
  monthly mean maximum, minimum and mean temperatures for the United States}.
\newblock {\em Journal of Climate and Applied Meteorology\/}~{\em 25\/}(2),
  145--160.

\bibitem[\protect\citeauthoryear{Menne, Durre, Vose, Gleason, and
  Houston}{Menne et~al.}{2012}]{menne2012overview}
Menne, M.~J., I.~Durre, R.~S. Vose, B.~E. Gleason, and T.~G. Houston (2012).
\newblock {An overview of the Global Historical Climatology Network-Daily
  database}.
\newblock {\em Journal of Atmospheric and Oceanic Technology\/}~{\em 29\/}(7),
  897--910.

\bibitem[\protect\citeauthoryear{Menne and Williams~Jr}{Menne and
  Williams~Jr}{2009}]{menne2009homogenization}
Menne, M.~J. and C.~N. Williams~Jr (2009).
\newblock Homogenization of temperature series via pairwise comparisons.
\newblock {\em Journal of Climate\/}~{\em 22\/}(7), 1700--1717.

\bibitem[\protect\citeauthoryear{Menne, Williams~Jr, and Vose}{Menne
  et~al.}{2009}]{menne2009us}
Menne, M.~J., C.~N. Williams~Jr, and R.~S. Vose (2009).
\newblock {The US Historical Climatology Network monthly temperature data,
  version 2}.
\newblock {\em Bulletin of the American Meteorological Society\/}~{\em
  90\/}(7), 993--1007.

\bibitem[\protect\citeauthoryear{Peterson, Easterling, Karl, Groisman,
  Nicholls, Plummer, Torok, Auer, Boehm, Gullett, et~al.}{Peterson
  et~al.}{1998}]{peterson1998homogeneity}
Peterson, T.~C., D.~R. Easterling, T.~R. Karl, P.~Groisman, N.~Nicholls,
  N.~Plummer, S.~Torok, I.~Auer, R.~Boehm, D.~Gullett, et~al. (1998).
\newblock Homogeneity adjustments of in situ atmospheric climate data: a
  review.
\newblock {\em International Journal of Climatology\/}~{\em 18\/}(13),
  1493--1517.

\bibitem[\protect\citeauthoryear{Quinonero-Candela, Rasmussen, and
  Williams}{Quinonero-Candela et~al.}{2007}]{quinonero2007approximation}
Quinonero-Candela, J., C.~E. Rasmussen, and C.~K. Williams (2007).
\newblock Approximation methods for gaussian process regression.
\newblock {\em Large-scale kernel machines\/}, 203--224.

\bibitem[\protect\citeauthoryear{Sherman}{Sherman}{2011}]{sherman2011spatial}
Sherman, M. (2011).
\newblock {\em Spatial statistics and spatio-temporal data: covariance
  functions and directional properties}.
\newblock John Wiley \& Sons.

\bibitem[\protect\citeauthoryear{Stein}{Stein}{2012}]{stein2012interpolation}
Stein, M.~L. (2012).
\newblock {\em Interpolation of spatial data: some theory for kriging}.
\newblock Springer Science \& Business Media.

\bibitem[\protect\citeauthoryear{Trewin}{Trewin}{2013}]{trewin2013daily}
Trewin, B. (2013).
\newblock {A daily homogenized temperature data set for Australia}.
\newblock {\em International Journal of Climatology\/}~{\em 33\/}(6),
  1510--1529.

\bibitem[\protect\citeauthoryear{Vincent, Wang, Milewska, Wan, Yang, and
  Swail}{Vincent et~al.}{2012}]{vincent2012second}
Vincent, L.~A., X.~L. Wang, E.~J. Milewska, H.~Wan, F.~Yang, and V.~Swail
  (2012).
\newblock {A second generation of homogenized Canadian monthly surface air
  temperature for climate trend analysis}.
\newblock {\em Journal of Geophysical Research: Atmospheres\/}~{\em
  117\/}(D18).

\end{thebibliography}

\begin{appendices}

\section{Derivation of the analytic posterior for toy example}
\label{sec:analytical_posterior}

In this appendix we derive and compute the conditional distribution \(\Fcond\) for the toy example of \autoref{sec:toy_example}.
We denote by \(f_i(\cdot)\) and \(F_i(\cdot)\) the prior probability distribution function and cumulative distribution function of \(X_i\), i.e. the normal PDF and CDF with means and variances given by \autoref{eq:toyspec}.
Let \(\pij\) be the probability that \(X_i\) is the minimum of \(X\) and \(X_j\) is its maximum.
We also define \(\pisum = \sum_{j=1}^{100} \pij\), the probability that \(X_i\) is the minimum,
and \(\psumj = \sum_{i=1}^{100} \pij\), the probability that \(X_j\) is the maximum.
The cumulative distribution function of \(X_i\) is then given by:
\begin{equation}
\Pr\del{X_i \leq x \mid \Xmax, \Xmin} =
    \begin{cases}
        0 &\text{if } x < \Xmin \,, \\
        1 &\text{if } x \geq \Xmax \,, \\
        \pxx{i}{\bullet} 
            + \del{1 \!-\! \pxx{i}{\bullet} \!-\! \pxx{\bullet}{i}}
            \sbr{\frac{F_i(x) - F_i(\Xmin) }
                 {F_i(\Xmax) - F_i(\Xmin) }
                } 
            &\text{otherwise.}\\
    \end{cases}
\end{equation}
Meanwhile, \(\pij\) is proportional to:
\begin{equation}
    f_i(\Xmin)
    f_j(\Xmax)
    \prod_{k \neq i,j}^{100}
    \del{F_k(\Xmax) - F_k(\Xmin)} \,,
\end{equation}
which we compute for all \(i,j\) and normalize
to obtain the \(100 \times 100\) matrix \(\Pr\) of probabilities of each pair of element occupying the extremes.
We sum over its rows and columns to obtain \(\psumj\) and \(\pisum\).
While this algorithm has cubic complexity in the dimensionality \(p\) of \(X\),
for \(p=100\), it only take seconds to compute the entries of \(\Pr\) and evaluate \(\Pr\del{X_i \leq x \mid \Xmax, \Xmin}\) over a range of \(x\).
\autoref{fig:toy_quantiles}(b) shows the analytical quantiles of \(\Fcond\) marginally for each \(X_i\).

\end{appendices}

\end{document}